\documentclass[useAMS,usenatbib]{mn2e}
\usepackage{epsfig, color}

\newcommand{\ud}{\mathop{\mathrm{{}d}}\mathopen{}}

\title[Mutual Interaction Between Pop III Stars and DM]
      {The Mutual Interaction Between Population III Stars and Self-Annihilating Dark Matter}

\author[A. Stacy, A. H. Pawlik, V. Bromm, and A. Loeb]
       {Athena Stacy$^{1,2}$\thanks{E-mail: athena.stacy@berkeley.edu}, Andreas H. Pawlik $^{3}$,  Volker Bromm$^{4}$, and Abraham Loeb$^{5}$\\
        $^{1}$University of California, Berkeley, CA 94720, USA  \\
        $^{2}$NASA Goddard Space Flight Center, Greenbelt, MD 20771, USA \\
        $^{3}$Max-Planck-Institut f\"{u}r Astrophysik, Karl-Schwarzschild-Str. 1, 85741 Garching, Germany \\
        $^{4}$Department of Astronomy and Texas Cosmology Center, University of Texas, Austin, TX 78712, USA\\
        $^{5}$Astronomy Department, Harvard University, 60 Garden Street, Cambridge, MA 02138, USA}

\pagerange{\pageref{firstpage}--\pageref{lastpage}}
\pubyear{2013}

\begin{document}

\maketitle
\topmargin-1cm

\label{firstpage}

\begin{abstract}
 We use cosmological simulations of high-redshift minihalos to investigate the effect of dark matter annihilation (DMA) on the collapse of primordial gas.  
We numerically investigate the evolution of the gas as it assembles in a Population~III stellar disk.
 %Our numerical study begins with a cosmological simulation initialized at $z=100$. 
 %We focus on the first star-forming region formed within the computational box to study how heating and ionization through DMA alters the gas evolution.   
 We find that when DMA effects are neglected, the disk undergoes multiple fragmentation events beginning at $\sim 500$ yr after the appearance of the first protostar.  
 On the other hand, DMA heating and ionization of the gas speeds the initial collapse of gas to protostellar densities
 and also affects the stability of the developing disk against fragmentation, depending on the DM distribution.
 %and the central gas is instead much more warm and stable to fragmentation than in the case without DMA.  
 We compare the evolution when we model the DM density with an analytical DM profile which remains centrally peaked, 
 and when we simulate the DM profile using N-body particles (the `live' DM halo).   When utilizing the analytical DM profile,  DMA suppresses disk fragmentation for $\sim 3500$ yr after the first protostar forms, in agreement with earlier work.  
However, when using a `live' DM halo, the central DM density peak is gradually flattened due to the mutual interaction between the DM and the rotating gaseous disk,  reducing the effects of DMA on the gas, 
and enabling secondary protostars of mass $\sim 1$ M$_{\odot}$ to be formed within $\sim$ 900 yr.  
These simulations demonstrate that DMA is ineffective in suppressing gas collapse and subsequent fragmentation, rendering the formation of long-lived dark stars unlikely.  However, DMA effects may still be significant in the early collapse and disk  formation phase of primordial gas evolution. 
%even when accounting for the warming and stabilizing effect of DMA, the gas motion nevertheless quickly reduces the central DM densities to values below which DMA can suppress the formation of small secondary protostars or allow for the existence of the previously proposed `dark stars.'
\end{abstract}


\section{Introduction}

The first stars, formed out of metal-free gas and also known as Population III (Pop III), initially emerged at $z \ga 20$ within the center of dark matter-dominated minihalos of mass $\sim 10^6$ M$_{\odot}$  
(e.g. \citealt{haimanetal1996,tegmarketal1997,yahs2003,loeb&furlanetto2012}).  
It is believed that those Pop III stars of sufficient mass then emitted the first reionizing photons, while a fraction of Pop III stars began the metal-enrichment of the intergalactic medium
 upon their supernova deaths
%, thereby beginning the process of  transforming the universe to the metal-enriched and reionized state observed today
(e.g. \citealt{barkana&loeb2001,bromm&larson2004,ciardi&ferrara2005,glover2005,byhm2009,loeb2010}).
%(e.g. \citealt{kitayamaetal2004,syahs2004,whalenetal2004,alvarezetal2006,johnsongreif&bromm2007}). 
 %(IGM;\citealt{madauferrara&rees2001,moriferrara&madau2002,brommyoshida&hernquist2003,wada&venkatesan2003,normanetal2004,tfs2007,greifetal2007,greifetal2010,wise&abel2008,wise&abel2012,maioetal2011}; recently reviewed in \citealt{karlssonetal2011}),
Early numerical work implied that the first stars formed in isolation and reached very high masses
($\ga 100$ M$_{\odot}$; e.g. \citealt{abeletal2002,brommetal2002,bromm&loeb2004,yoh2008}).
However, more recent simulations have shown that primordial gas undergoes multiple fragmentation,
forming binary and multiple systems,
(e.g. \nocite{clarketal2008,clarketal2011a, clarketal2011b} Clark et al 2008, 2011a, 2011b;  
\citealt{turketal2009,stacyetal2010,greifetal2011,greifetal2012,bromm2013}). 
Simulations which include modelling of feedback from the growing protostar have furthermore shown that ionizing radiation will dissipate away the accretion disk once the star has reached $\sim$ 40 M$_{\odot}$ (e.g. \citealt{hosokawaetal2011}; \nocite{stacyetal2012a} Stacy et al. 2012a).  
Though these star-formation processes go on within minihalos whose mass is composed of nearly ten times more dark matter (DM) than baryons, this picture of Pop III stars does not account for the influence of DM beyond providing a gravitational well for the initial collapse of primordial gas.
 %40~M$_{\odot}$~$<$~$M_{*}$~$<$~140~M$_{\odot}$ or  $M_{*}$~$>$~260~M$_{\odot}$
 %140~M$_{\odot}$~$<$~$M_{*}$~$<$~260~M$_{\odot}$ 
  %(PISNe; \citealt{heger&woosley2002})
   %(\citealt{chatz&wheeler2012}).  
%15~M$_{\odot}$~$<$~$M_{*}$~$<$~40 M$_{\odot}$ (e.g. \citealt{nomotoetal2003}).  
%\cite{mckee&tan2008} 
%100~M$_{\odot}$.  
%(e.g. \citealt{turketal2009,stacyetal2010}).  
%(Clark et al. 2011b, \nocite{clarketal2011b} \citealt{greifetal2011}),  
%(\citealt{smithetal2011}).  

However, the nature of the DM particle remains unknown.  One intriguing possibility is that DM is made up of weakly interacting massive particles (WIMPS, see \citealt{bertoneetal2005} for a review).  If this is the case, then Pop III star formation may significantly differ from the above description.  This is because WIMPs are self-annihilating, and such DM annihilation (DMA) releases a cascade of photons that will potentially alter the thermal and chemical state of the primordial gas if the DM density and annihilation rate are sufficiently large.  
DMA may first become significant during the initial collapse of gas within a minihalo.  This is due to the growing density of DM in the center of the minihalo as it responds to the increasing potential well of the gas, generally termed `gravitational contraction.'  
DMA may subsequently influence the characteristics of the resulting protostars as well as the evolution of the protostellar disk.
For instance, several studies (e.g. \citealt{spolyaretal2008, freeseetal2008,natarajanetal2009}) find that gravitational contraction leads to sufficient DM annihilation to halt the collapse of the primordial cloud before 
%a hydrostatic object has formed, 
hydrogen is ignited in its center,
leading instead to the formation of what has been termed a `dark star', a giant ($\sim$ 1 AU) star powered by DM annihilation instead of nuclear burning.  

Because of the extended nature of these objects, their effective temperatures are too low to emit ionizing radiation.  The DM supply to a dark star may continue through gravitational accretion, and depending on how long this process is sustained, the lack of protostellar feedback may allow for a prolonged period of gas inflow before the dark star phase ends.  Afterwards the star begins contraction to the main sequence (MS), and radiative feedback shuts off mass inflow onto the star (e.g. \citealt{spolyaretal2009}). 
\cite{ioccoetal2008} find that the dark star phase is short-lived, only $\sim 10^4$ yr,
though they did not follow the gradual accretion of gas onto the stars over time.  
More detailed one-dimensional work (\citealt{ripamontietal2010}) even finds that gravitational accretion of DM during the initial cloud contraction 
%will not halt the gas collapse  
does not halt or significantly alter the initial gas collapse up to densities of $\ga 10^{14}$ cm $^{-3}$.  They interpret this to mean a dark star does not form, though \cite{gondoloetal2013} point out that dark star formation actually occurs at higher densities of $\sim 10^{17}$ cm $^{-3}$ which \cite{ripamontietal2010} do not resolve.  
%Thus, the initial mass of the first stars will not be significantly affected.  However, other work reaches different conclusions (e.g. \citealt{freeseetal2008}).

At later stages, Pop III protostars can nevertheless continue gathering DM through continued gravitational accretion as the protostar's mass and potential well grows.  Even a contracted MS star may gain DM particles through `scattering accretion,' a process in which a WIMP scatters off the gas of the star,
loses kinetic energy in the scattering event(s), and sinks to the center of the star (e.g.Freese et al. 2008b\nocite{freeseetal2008b},  \citealt{iocco2008}).
%and become gravitationally bound to it.  
If the scattering cross section between WIMPs and baryons is large enough, and the resulting capture rate of DM by Pop III stars sufficiently high, this would prolong the lifetimes of Pop III stars.  This is because hydrogen will burn at a reduced rate while DM annihilation helps to support the star.  Recent work by \cite{sivertsson&gondolo2011}, however, find that this phase of scattering accretion will be very short-lived, $\la$ 10$^5$ yr, much shorter than the lifetime of the star.

Most recently, the smoothed particle hydrodynamics (SPH)
simulations of \nocite{stacyetal2012} Stacy et al. (2012b) included both gaseous (SPH) and N-body particles to resolve the three-dimensional evolution of both primordial 
gas and DM on scales as small as $\sim$ 50 AU. This study found that gravitational interaction between the DM and the fragmenting protostellar disk causes the DM densities to rapidly decline.  This drives the DM density below the threshold necessary for DMA rates to significantly influence the gas, indicating that DMA effects will not have long-lasting effects on primordial disk evolution.  
However, the \nocite{stacyetal2012}  Stacy et al. (2012b) simulations did not include DMA rates in following the chemothermal evolution of the gas, and they did not resolve the inner disk on scales of $< 50$ AU.  

%\cite{smithetal2012} addressed this by performing three-dimensional SPH simulations which followed the collapse of primordial minihalo gas under DMA effects.  
\cite{smithetal2012}  addressed the effects of DMA by performing three-dimensional SPH simulations which followed the collapse of primordial gas in minihalos.
In agreement with \cite{ripamontietal2010}, they found that DMA does not halt gas collapse to near-stellar densities.  
Their work furthermore indicated that, 
following the formation of the first protostar, DMA modifies the evolution of the star-forming disk by stabilizing it against fragmentation.  Instead of a multiple system which undergoes protostellar mergers and ejections, any fragmentation leads to at most a wide binary.  
\cite{smithetal2012} thus called into question the 
above-mentioned studies which found that primordial disks will typically form several protostellar fragments. 
The stabilization of the disk may even prolong the effects of DMA, as a smooth disk may not disrupt the central DM densities as quickly a stellar multiple system.  
%result of \cite{stacyetal2012}.  
However, in \cite{smithetal2012} the DM profile was static.  The DM density was determined with an analytic prescription dependent upon distance from the gas density peak, and the peak of the DM density profile was fixed to align with the position of the first-formed protostar.  
Their one-dimensional spherically symmetric DM profile was therefore unable to follow the gravitational interaction between the gas and the DM.
%, which other studies have found can become particularly important after the development of a rotating protostellar disk (\citealt{stacyetal2012}). 

In this work we avoid the approximations discussed above.  We perform simulations which include both DMA heating and ionization of the gas, as well as the mutual gravitational interaction between the gas and the DM.  We resolve both gas and DM down to scales of $\sim$ 5 AU, and include a prescription for DMA rates based upon the local DM density.  Our simulations therefore provide the most physically realistic representation of Pop III star formation under the influence of DMA. 
As in \cite{smithetal2012} and \cite{ripamontietal2010}, we find that DMA does not halt gas collapse up to near protostellar densities.  
We next employ the sink particle method to examine the mutual evolution of the primordial disk and central DM density.  
However, we do not address the question of whether the equilibrium object that forms on sub-sink scales is a `dark star' or normal protostar, leaving this for future studies.
In Section 2 we discuss our numerical methodology, while in Section 3 we discuss our handling of DM.  We present our results in Section 4 and conclude in Section 5.
\\
\\
\begin{table*}
\centering
\begin{tabular}{ccccc}
\hline
Name & DM profile type & $m_x$ (GeV/$c^2$) & $n_{\rm frag}$ & $M_{*, \rm tot}$ [M$_{\odot}$]\\ 
\hline
no-DMA       & no DMA           & N/A              & 3 & 7\\
DMA-A1 & analytic, Smith et al. (2012) &  100 & N/A & N/A\\
DMA-A2 & analytic, $\rho \propto r^{-2}$ & 100 & 4 & 14\\
DMA-L2 & live,  $\rho \propto r^{-2}$ &  100 & 7 & 16\\
\hline
\end{tabular}
\caption{Summary of simulations described in this work. 
Total number of sinks formed is $n_{\rm frag}$, and $M_{\rm *, tot}$ is total sink mass accreted throughout the simulation.
Run DMA-A1 was stopped just before the formation of the first sink.
}
\label{tab1}
\end{table*}
\\
\\

\section{Numerical Methodology}

\subsection{Initial Setup}
We carry out our investigation using {\sc gadget-2,} a widely-tested three-dimensional N-body and SPH code (\citealt{springel2005}). 
We begin with a 140 $h^{-1}$ kpc (comoving) box containing 128$^3$ SPH gas particles and 128$^3$ DM particles.  The simulation is initialized at $z=100$.   Positions and velocities are assigned to the particles in accordance with a 
$\Lambda$ cold dark matter cosmology with $\Omega_{\Lambda}=0.7$, $\Omega_{\rm M}=0.3$, $\Omega_{\rm B}=0.04$, $\sigma_8=0.9$, and $h=0.7$.  The gas and DM evolution is followed  until the first minihalo forms and its central gas density reaches 
$10^4$ cm$^{-3}$. 

%Once the site of the first minihalo is determined, 
In the second step, the simulation is performed at higher resolution, starting again at $z=100$.  We increase the resolution using a hierarchical zoom-in procedure 
(e.g.  \citealt{navarro&white1994,tormenetal1997,gaoetal2005}) in which four nested refinement boxes are placed within the cosmological box.  
The refinement boxes are centered on the site where the minihalo will eventually form, as determined by the original unrefined simulation.  
The four refinement levels have lengths of 40, 35, 30, and 20 $h^{-1}$ kpc (comoving), so that the most highly refined level encompasses all the mass that will later be incorporated into the minihalo.  
Within each refinement level, each particle from the lower level is replaced with eight `child' particles, so that in the most refined region the parent particle is replaced by 4096 child particles.
The most refined SPH particles have mass $m_{\rm SPH} = 5 \times 10^{-3}$ M$_{\odot}$, and
the resolution mass of the refined simulation is 
$M_{\rm res}\simeq 1.5 N_{\rm neigh} m_{\rm SPH}\la  0.3 $M$_{\odot}$, where $N_{\rm neigh}\simeq 40$ is the typical number of particles in the SPH smoothing kernel (e.g. \citealt{bate&burkert1997}).

\begin{figure*}
\includegraphics[width=.8\textwidth]{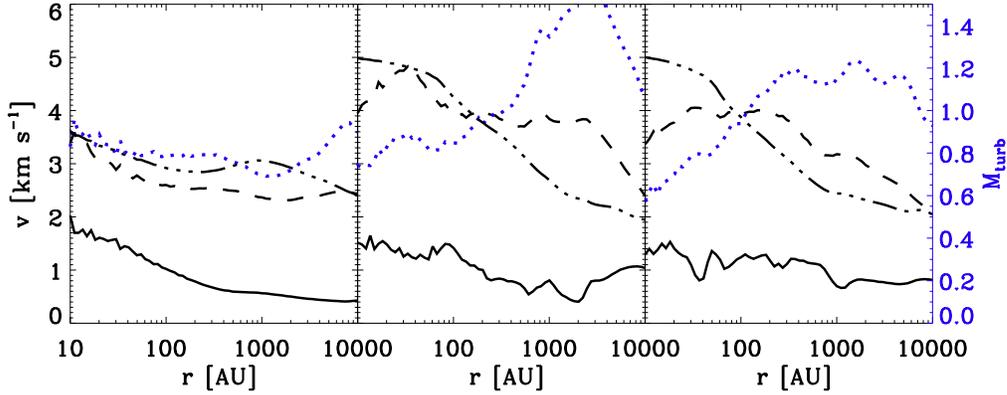}
\caption{
Gas velocity profiles of the simulations at the point just prior to sink formation, when the maximum density has reached 10$^{16}$ cm$^{-3}$.
{\it Left:} No DMA effect.
{\it Center:}  DMA effects for Smith et al. profile (DMA-A1).
{\it Right:}  DMA effects for $r^{-2}$ profile (DMA-A2).
Solid lines: the magnitude of the radial velocity $|v_{\rm rad}|$.  The sign of $v_{\rm rad}$ is negative and gas is inflowing.  
Dashed lines: rotational velocity $v_{\rm rot}$.
%Dash-dotted lines: free-fall velocity $v_{\rm ff}$, as determined by enclosed mass at the given radius.
Dash-dotted lines: sound speed.
Blue dotted lines: turbulent Mach number $M_{\rm turb}$, which follows the scaling shown on the right-hand y-axis.  
When DMA effects are included, the inflow within the central 10,000 AU exhibits slightly greater radial velocities at $r>1000$ AU as well as enhanced turbulence and rotational motion. 
%as well as a greater enclosed mass, as depicted by the enchanced $v_{\rm ff}$.  
In agreement with Smith et al. (2012), the sound speed in the central 1000 AU is larger when compared with the no-DM case but somewhat smaller beyond 1000 AU.
}
\label{velprof}
\end{figure*}

\subsection{Cut-Out Technique}
%The computational timesteps of the calculation become smaller as the gas density grows.  
To increase the computational efficiency of the simulation,  once the gas has reached number densities of $10^{12}$ cm$^{-3}$ we employ a `cut-out' technique in which we remove all gas and DM outside the 10 pc cube that is centered on the densest gas particle.  At this stage, the central star-forming gas is gravitationally bound and  under minimal influence from the mass of the outer minihalo and more distant regions of the cosmological box.  The total mass of the cut-out is 3500 M$_{\odot}$, and the minimum density is $\sim 10^2$ cm$^{-3}$.  

The gas at the cut-out edge has a free-fall time of $\sim 10^7$ yr and will undergo little evolution over the next $\sim$ 10,000 yr followed in the simulation.  We note that this cut-out technique causes the propagation of a rarefaction wave which originates from the cut-out edge due to the vacuum boundary condition.  However, this wave will only travel a distance of $c_{\rm s} \, t$, where $c_{\rm s}$ is the gas sound speed ($\sim$ 2 km s$^{-1}$), and the time $t$ is 10,000 yr.  This corresponds to a negligible distance of $\sim$ 10$^{-2}$ pc (2000 AU) from the cut-out edge, three orders of magnitude smaller than the 10 pc box size.

\subsection{Particle Splitting}
At the same time that we cut out the central 10 pc of the cosmological box, we further increase the mass resolution so that collapse to densities of 10$^{16}$ cm$^{-3}$ can be properly followed.  
We thus replace each SPH particle with 8 child particles, each of which is placed randomly within the smoothing kernel of the parent particle.  The mass of the parent particle is then evenly divided amongst the child particles.  Each of these particles inherits the same chemical abundances, velocity, and entropy as the parent particle (see, e.g., \citealt{bromm&loeb2003}, \nocite{clarketal2011b} Clark et al. 2011b).  This ensures conservation of mass, internal energy, and linear momentum, but not of the center of mass. 
%and the angular momentum.  
Each SPH particle in the new cut-out simulation has a mass of $m_{\rm sph}=6\times10^{-4}$ M$_{\odot}$, and the new resolution mass is $M_{\rm res} \simeq 0.03$ M$_{\odot}$.  
 $M_{\rm res}$ is close to the mass of the pressure-supported atomic core (10$^{-2}$ M$_{\odot}$) which develops once the opacity limit is reached  (\citealt{yoh2008}), defining the point at which the protostar has first formed.

\subsection{Chemistry, Heating, and Cooling}
We utilize the same chemistry and thermal network as described in detail by \cite{greifetal2009} and used in \nocite{stacyetal2012}  Stacy et al. (2012b).   In short, the code follows the abundance evolution of  
H, H$^{+}$, H$^{-}$, H$_{2}$, H$_{2}^{+}$, He, He$^{+}$, He$^{++}$, and e$^{-}$, as well as the three deuterium species D, D$^{+}$, and HD.  
All relevant cooling mechanisms, including H$_2$ collisions with  H and He as well as other H$_2$ molecules, are modelled in our thermal network.  
The thermal network also includes cooling through  
H$_2$ collisions with protons and electrons, H and He collisional excitation and ionization, recombination, bremsstrahlung, and inverse Compton scattering.  

We must also account for further H$_2$ processes to properly model gas evolution to high densities.  This  includes three-body H$_2$ formation and the concomitant H$_2$ formation heating, which become important at  $n \ga 10^8$ cm$^{-3}$, where $n$ refers to the total number density for all species. 
The three-body H$_2$ formation rate is uncertain.  Previously published values vary by nearly an order of magnitude, such that employing a different rate coefficient in a simulation may yield significant variation in the gas properties at these high densities (\citealt{turketal2011}).  We note this caveat, and we use the rate presented in \cite{pallaetal1983} because it is intermediate among the range of suggested values.

When $n \ga 10^9$ cm$^{-3}$, cooling through H$_2$ ro-vibrational lines becomes less effective as these lines grow optically thick.  We model this effect by multiplying the H$_2$ line cooling rate by an escape probability factor:
\begin{equation}
\beta_{\rm esc} = \beta_{\rm esc} (\tau_{\rm H_2})= \frac{1-{\rm e}^{-\tau_{\rm H_2}}}{\tau_{\rm H_2}} 
\end{equation}
(Clark et al. 2011a), 
where $\tau_{\rm H_2}$ is determined using the Sobolev approximation
(see also \citealt{yoshidaetal2006,greifetal2011, hirano&yoshida2013} for further details).  

The most important new process utilized in the thermal network was H$_2$ collision-induced emission (CIE) cooling, in which an H$_2$-H$_2$ collision (or less importantly an H$_2$-H or  H$_2$-He collision)
%does not lead the H$_2$ to inhabit an excited state as a single H$_2$ molecule, but instead 
briefly creates a `supermolecule' with a non-zero electric dipole field and a high probability of emitting a photon (\citealt{ripamonti&abel2004}).  CIE cooling becomes significant at densities of $n \ga 10^{14}$ cm$^{-3}$ (\citealt{frommhold1994}), at which point it is more effective than  H$_2$ line cooling.  
%Inclusion of CIE was necessary to properly model the collapse of gas to the maximum densities of 10$^{16}$ cm$^{-3}$.  
%As described in \cite{greifetal2011}, 
Similar to the above-mentioned handling of H$_2$ line opacity, the reduction of the CIE cooling rate due to the effects of continuum opacity is approximated through the following prescription (\citealt{ripamontietal2002,ripamonti&abel2004}):
\begin{equation}
%\Lambda_{\rm CIE, thick}=\Lambda_{\rm CIE, thin}\,{\rm min}\left(\frac{1-e^{-\tau_{\rm CIE}}}{\tau_{\rm CIE}},1\right)\mbox{\ ,}
\Lambda_{\rm CIE, thick}=\Lambda_{\rm CIE, thin}\,{\rm min}\left[ \beta_{\rm esc} (\tau_{\rm CIE}), 1 \right]\mbox{\ ,}
\end{equation}

\noindent where $\Lambda_{\rm CIE, thin}$ is the CIE cooling rate in the optically thin limit, and $\Lambda_{\rm CIE, thick}$ is the rate in optically thick conditions.
Unlike the optically thick H$_2$ line cooling, however, the optical depth is estimated with a fitting formula instead of the Sobolev approximation:
\begin{equation}
\tau_{\rm CIE}=\left(\frac{n_{\rm H_2}}{7\times 10^{15}\,{\rm cm}^{-3}}\right)^{2.8}\mbox{\ .}
\end{equation}

When including DMA heating and ionization (Section 3.1), we found an unphysically large cooling rate due to conversion of ortho-H$_2$ to para-H$_2$ in the $J = 0 \rightarrow 1$ transition.  This is because we made the usual assumption of a constant ortho-para ratio of 3:1 (\citealt{glover&abel2008}).  The cooling/heating rate $\Lambda_{\rm OP}$ is given by 
\begin{equation}
%\Lambda_{\rm OP} = 4.76 \times 10^{-24} n_{\rm H^+} n_{\rm H_2} \left[ 9 \, {\rm exp} \left( \frac{-170.5 \rm K}{T} \right)x_{\rm p} - x_{\rm o} \right] {\rm  \frac{erg} {cm^{3} s}} 
\Lambda_{\rm OP} = R_{\rm OP} \, n_{\rm H^+} n_{\rm H_2} \left[ 9.0 \, {\rm exp} \left( \frac{-170.5 \rm K}{T} \right)x_{\rm p} - x_{\rm o} \right]
\end{equation}

% Ortho-para conversion heating / cooling
%        rates(2) = rates(2) + 4.76d-24 * ynhp * ynh2 *
%     $             (cl67 * 0.25d0 - 0.75d0)
\noindent (\citealt{glover&abel2008}), where  
$R_{\rm OP} = 4.76 \times 10^{-24} \, {\rm erg \, cm^{3} \, s^{-1}}$,
$n_{\rm H^+} $ and $n_{\rm H_2} $ are the densities of ionized hydrogen and molecular hydrogen,
and $x_{\rm p}$ and $x_{\rm o}$ are the fractions of H$_2$ that are in the para and ortho states, respectively.  
$\Lambda_{\rm OP}$  is given in units of energy per time per volume.  
$\Lambda_{\rm OP}$ does not dominate the cooling and heating of the gas in the canonical case without DMA, when the $n_{\rm H^+} n_{\rm H_2}$ factor remains low.   
However, with DMA effects the above estimate for $\Lambda_{\rm OP}$ becomes several orders of magnitude larger and causes the high-density ($n \ga 10^7$ cm$^{-3}$) gas to cool to nearly the CMB floor.  
This is due the great enhancement of the H$_2$ and e$^-$ fraction through DMA effects.

\begin{figure}
\includegraphics[width=.45\textwidth]{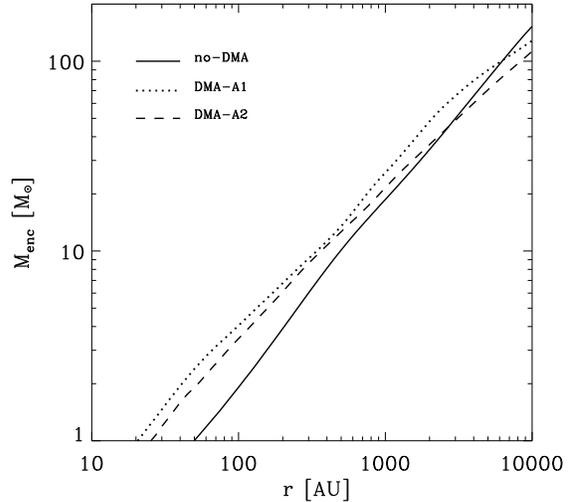}
\caption{
Enclosed mass $M_{\rm enc}$ with respect to radius for three cases: no-DMA (solid line), DMA-A1 (dotted line), and DMA-A2 (dashed line).  
In each case, profiles are measured just before initial sink formation, and the distance $r$ is measured with respect to the densest gas particle.
The effects of DMA lead to greater $M_{\rm enc}$ within the central few thousand AU.
}
\label{menc_prof}
\end{figure}

%To address this problem, 
In most cases assuming a 3:1 ortho-para ratio gives a very close approximation to self-consistently calculating the ratio from H$_2$ level populations (\citealt{glover&abel2008}), but this does not hold when including DMA effects.
For an improved approximation, in all cases we set $\Lambda_{\rm OP} = 0$ after gas reached densities greater than 10$^5$ cm$^{-3}$.  
%???
This is more physically correct given that the ortho-para ratio should be in equilibrium at the densities we consider, while  $\Lambda_{\rm OP}$ will be active only in cases of non-equilibrium.  
As discussed in \cite{glover&abel2008}, in the case of thermodynamic equilibrium, ortho- to para-H$_2$ conversions will balance with the para- to ortho-H$_2$ conversions such that there will be no net effect on temperature.  This equilibrium will be reached especially quickly under the  enhanced H$^+$ fractions under DMA effects (see equation 24 of \citealt{glover&abel2008}).  For instance, note that at densities of $n=10^{10}$ cm$^{-3}$ and H$^+$ fractions of $10^{-5}$, the ortho-para ratio will reach equilibrium in $\ga 3 \times 10^{-3}$ yr, much more quickly than the H$_2$ formation time of $\sim$ 10,000 yr or the dynamical time of $\sim$ 1000 yr.
Thus, when including DMA effects, removing the $\Lambda_{\rm OP}$ term is necessary for the gas to reproduce the thermal evolution seen in \cite{smithetal2012} and to avoid cooling to the CMB.
As will be seen in Section 3, this is in contrast to the no-DMA case in which inclusion or removal of $\Lambda_{\rm OP}$ has little effect on the thermal and chemical evolution of the gas, such that either way the evolution remains very similar to results found in other simulations.

\begin{figure*}
\includegraphics[width=.8\textwidth]{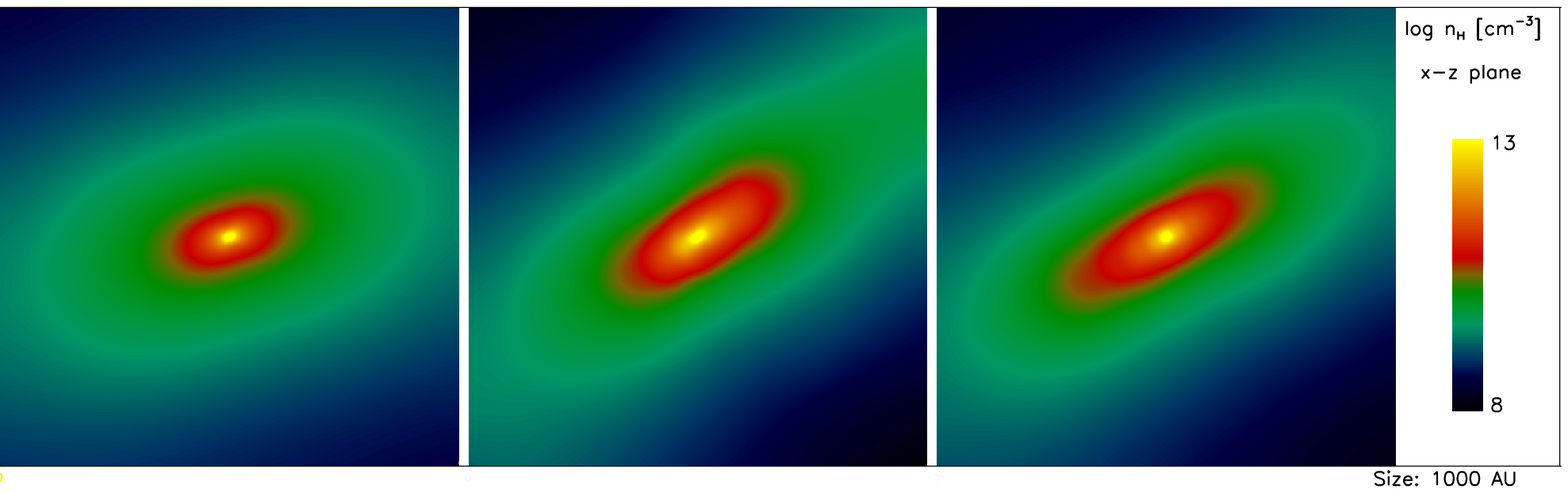}
\includegraphics[width=.8\textwidth]{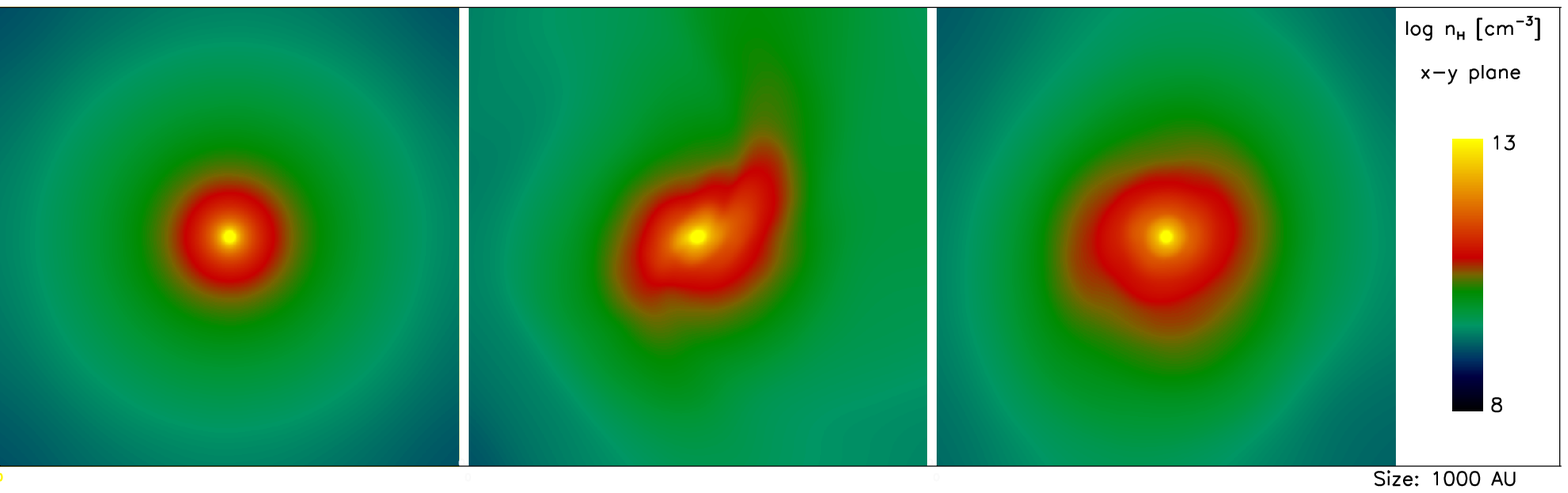}
 \caption{
Gas densities in the central 1000 AU.  Top and bottom rows show the same regions but oriented along orthogonal planes.
{\it Left:} No DMA effects (no-DMA).
{\it Center:} DMA effects for analytic halo with Smith et al. 2012 profile (DMA-A1).
{\it Right:}  DMA effects for analytic halo with $r^{-2}$ profile (DMA-A2).
Note that DMA leads to more flattened and extended disky structures on these small scales.
 }
\label{morph}
\end{figure*}

\subsection{Sink Particle Method}
To become a sink, an SPH particle must reach a number density of $n_{\rm max} = 10^{16}$ cm$^{-3}$, and gas within its smoothing length must have a negative velocity divergence.  The sink then grows in mass by accreting surrounding particles within its inner accretion radius $r_{\rm acc}$.  We set $r_{\rm acc}$ equal to several times the baryonic resolution length of the simulation such that
 $r_{\rm acc} = 6\,L_{\rm res} \simeq 6$ AU, where

\begin{equation}
L_{\rm res}\simeq \left(\frac{M_{\rm res}}{\rho_{\rm max}}\right)^{1/3} \mbox{\ .}
\end{equation}

\noindent The sink accretes gas particles within $r_{\rm acc}$ unless they are rotationally supported against infall onto the sink.  This is determined by checking that the particle  satisfies  $j_{\rm SPH} < j_{\rm cent}$, 
where $j_{\rm SPH} = v_{\rm rot} d$ is the angular momentum of the gas particle, $j_{\rm cent} = \sqrt{G M_{\rm sink} r_{\rm acc}}$ is the angular momentum required for centrifugal support, and $v_{\rm rot}$ and $d$ are the
rotational velocity and distance of the particle relative to the sink.  SPH particles that satisfy these criteria are removed from the simulation, and their mass is added to that of the sink.  
In practice, we have found that including the angular momentum criterion does not significantly affect the sink accretion rates by more than a few percent (\citealt{stacyetal2010, stacyetal2011}).  
%Most particles at close distances to the sink have highly circular orbits, while only particles with elliptical orbits will pass the distance criterion but fail the angular momentum criterion.
Most particles that come within $d < r_{\rm acc}$ have highly circular orbits with $j_{\rm sph}  < j_{\rm cent}$, while only particles with elliptical orbits that pass the distance criterion will fail the angular momentum criterion.

SPH particles that approach inside an inner accretion radius of $r_{\rm acc,in} = (1/2) r_{\rm acc}  \simeq 3$ AU are always accreted by the sink, regardless of their angular momentum.  
This prevents motion of particles that come this close to a sink from being dominated by numerical N-body effects, since in physical reality the hydrodynamic forces of unresolved gas and disks on these scales will counter purely gravitational effects.
When the sink first forms, it immediately accretes most of the particles within its smoothing length, so its initial mass is a few times the resolution mass of the simulation, 
$M_{\rm res} \simeq 3 \times 10^{-2}$ M$_{\odot}$.  Its position and velocity are set to the mass-weighted average of that of the accreted particles.  
These same accretion criteria also determine whether two sinks may be merged.  However, no sink mergers occur in the cases we present here.

%These same accretion criteria also determine whether two sinks may be merged.  However, we note that modifications to the sink merging algorithm can significantly alter the sink accretion history (\citealt{greifetal2011}). 
%Recent work by \cite{greifetal2012} has resolved sub-protostellar scales (0.05 R$_{\odot}$) of primordial star-forming gas, tracking the merger rate of protostars by  following their interactions without using sinks.  They find that approximately half of the secondary protostars formed will indeed migrate towards and merge with the initial protostar.  
%Our merging algorithm leads to a similar fraction of secondary sinks merging with the main sink, and thus reflects well what occurs on sub-sink scales.  

After accretion of a new gas particle, the sink's position and velocity are updated to the mass-weighted average of the sink and the accreted particle or secondary sink.   The sink is given a constant density of $10^{16}$ cm$^{-3}$ and temperature of 2000 K, the typical temperature for gas at this density.  
The sink thus exerts a pressure on the surrounding particles.  This prevents the formation of an artificial pressure vacuum around its accretion radius (see \citealt{bateetal1995,brommetal2002,marteletal2006}).  

Our handling of sinks is very similar to that of \cite{smithetal2012}.  They employed the same $n_{\rm max}$ criterion for sink formation, though they also imposed additional criteria that within the initial sink radius gravitational energy dominates over thermal and rotational energy, while the total energy of the particles must be negative.  They used similar $r_{\rm acc}$ and $r_{\rm acc,in}$ radii of 6 and 4 AU.  Their further requirement that accreted SPH particles be gravitationally bound to the sink is equivalent to our angular momentum criterion.  As will be seen in Section 3, the small differences in our sink prescriptions did not lead to significantly different results in comparable cases.

%Through the use of sink particles we thus avoid the problem of ever-decreasing computational timesteps due to ever increasing densities.  This increases the time period over which the surrounding star-forming gas can be evolved to approximately thousands of free-fall times.  The simulation can thus follow the further disk formation and fragmentation of the gas while still resolving the protostellar accretion rate on very small scales.  

\section{Dark Matter Prescription}

\subsection{Dark Matter Annihilation Rates}

We compute the contribution of DMA energy to the thermal and chemical evolution of the halo gas using the prescription described in \cite{smithetal2012}.  We briefly outline their method below.  
We first assume the standard DMA cross section of $\langle \sigma v \rangle = 3 \times 10^{-26}$ cm$^3$s$^{-1}$ (e.g. \citealt{spolyaretal2008}).
This is the canonical cross section derived from the `thermal freeze-out' scenario, and it is largely independent of the DM particle mass  $m_x$ (\citealt{jungmanetal1996}), though observational upper limits generally allow for a larger $\langle \sigma v \rangle$ given a greater $m_x$ (e.g. \citealt{ando&komatsu2013}).
For a DM density $\rho_{x}$ and mass $m_x$, the DM number density is $n_x = \rho_x/m_x$.  From this we can determine the number of DMA interactions per unit volume per unit time:

\begin{equation}
\dot{n}_{\rm DMA} = \frac{1}{2}n_x^2 \langle \sigma v \rangle \mbox{.}
\end{equation}

\noindent As in \cite{smithetal2012}, we use the results of \cite{valdes&ferrara2008} and set

\begin{equation}
f_h = 1 - 0.874\left(1 - x_e^{0.4052} \right) 
\end{equation}

\noindent and

\begin{equation}
f_i = 0.384\left(1 - x_e^{0.542} \right)^{1.1952} \mbox{,}
\end{equation}

\noindent where $x_e$ is the electron abundance fraction, $f_h$ is the fraction of DMA energy deposited into heat, and $f_i$ is the fraction that contributes to ionizations and dissociations.  
%As in \cite{ripamontietal2007} and \cite{smithetal2012}, 
Finally, we take the DMA heating rate per unit volume per unit time to be

\begin{equation}
\Gamma_{\rm DMA} =  2 \, f_a \, \dot{n}_{\rm DMA} \, \left(1 - {\rm e}^{-\tau_x} \right) \, f_h \, m_x \, c^2\mbox{,} 
\end{equation}

\noindent where $\tau_x$ is the optical depth of the gas to annihilation products, and $f_a = 2/3$ is the fraction of the total DMA energy that affects the gas, while the remaining 1/3 escapes as neutrinos.    
%Similarly, the rate of DMA energy injection per unit volume due to ionizations and dissociations is 
Similarly, a portion of the injected DMA energy contributes to ionization and dissociation of the gas.  This occurs at a rate per unit volume of

\begin{equation}
Q_{\rm DMA} = 2 \, f_a \, \dot{n}_{\rm DMA} \, \left(1 - {\rm e}^{-\tau_x} \right) \, f_i \, m_x \, c^2 / \epsilon_i \mbox{,}
\end{equation}

\noindent where $\epsilon_i$ is the threshold energy for the relevant ionization or dissociation reaction.
Ionization rates are applied to H, D, He, and He$^+$, and dissociation rates are applied to H$_2$, HD, and H$_2^+$.
For each of our simulations which include DMA effects, we take a fiducial value of $m_x = 100 {\rm GeV}/ {\it c^2}$.
%Note that $Q_{\rm DMA} \propto (1/m_x)$.  This is because $Q_{\rm DMA} \propto \dot{n}_{\rm DMA} \, m_x \propto n_x^2 \, m_x \propto (\rho_x/m_x)^2 \, m_ x$.  A factor of ten decrease in the DM particle mass thus leads to a factor of ten increase of the DMA rates.

As in \cite{smithetal2012}, we estimate $\tau_x$ based on the gas particle's distance $r$ from the central gas density peak
and the size of the central DM core $r_c$, which varies as the baryonic and DM density profiles evolve 
(see the following section):
\begin{equation}
\tau_x \equiv \kappa \Sigma(r) = \left \{ \begin{array}{lr}
\kappa \rho(r) (2r_{c} - r) & r \leq r_{c} \\
\kappa \rho(r) r & r > r_{c}
\end{array} 
\right.
\end{equation}

\noindent where we employ a constant gas opacity of $\kappa = 0.01$ cm${^2}$ g$^{-1}$.  This formulation for optical depth makes the simplifying assumption that the gas density is constant within the DM core and declines as $\rho \propto r^{-2}$ at larger radii.

\begin{figure*}
\includegraphics[width=.8\textwidth]{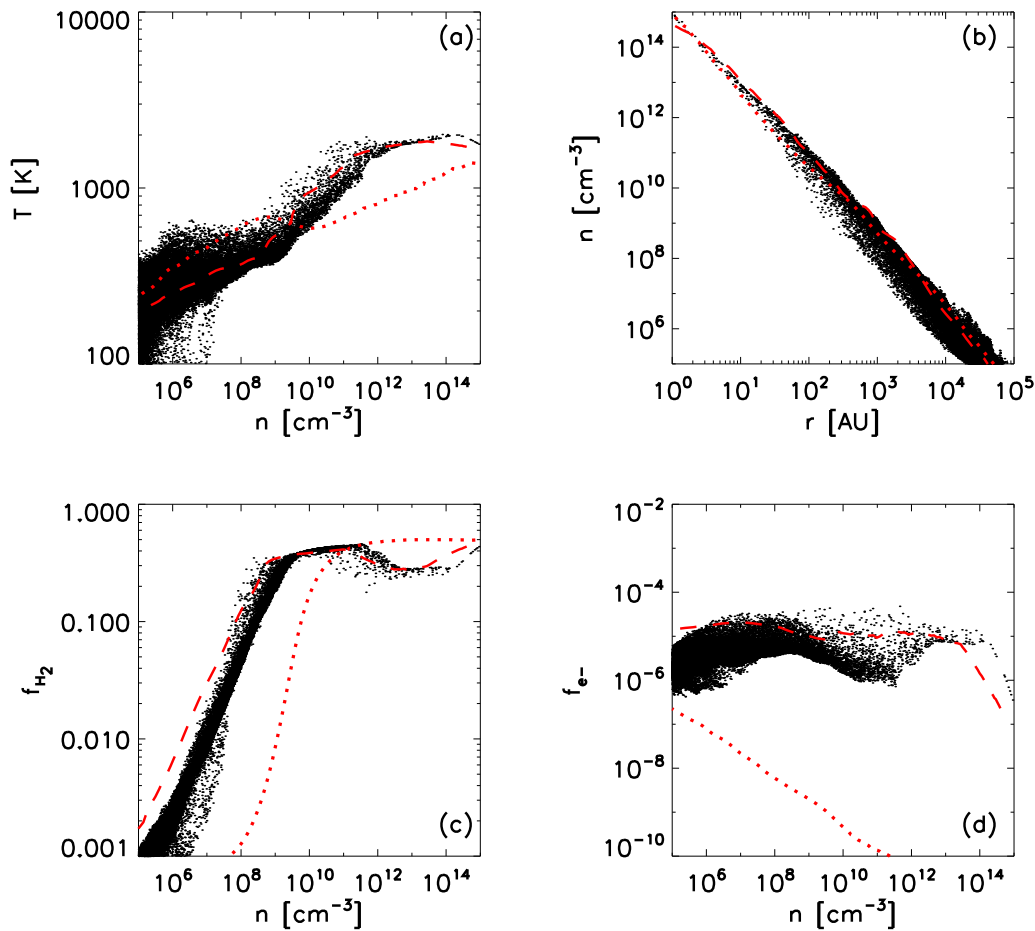}
 \caption{
Thermal and chemical structure of the gas in the case of the analytic DM halo DMA-A2, shown at the point of initial sink formation ($t=0$).
For comparison, the dashed line shows case DMA-A1, which utilized the DM density profile of Smith et al. (2012).  Note the similarity in chemothermal evolution for both of the analytic DM profiles.
The dotted line shows the case without DMA effects (no-DMA), which is very similar to previous results found by other authors (e.g. Clark et al 2011b).  Note that $f_{\rm H_2}=0.5$ corresponds to fully molecular gas.
 }
\label{Tvsnh}
\end{figure*}

\subsection{DM Density Profiles}

\subsubsection{Analytic Profiles}

The main input into the DMA heating and ionization rates is the DM density $\rho_x$.  In our simulation `DMA-A1', we employ the same $\rho_x$ profile as used in Smith et al. (2012; see also references therein).  The density at the edge of the core is   
\begin{equation}\label{dmpeak}
\rho_{xc} \approx 5 n_{\rm c}^{0.81} \:  {\rm \frac{GeV}  {\it c^2}  cm^{-3}}  ,
\end{equation} 
where $n_{\rm c}$ is the peak gas number density within the core, such that $n_{\rm c}$  evolves as the gas condenses and eventually reaches a maximum of 10$^{16}$ cm$^{-3}$.  At distances beyond the core ($r > r_{\rm c}$), the DM density declines as
\begin{equation}\label{dmprofile}
\rho_{x} = 5 \times 10^4 \left(\frac{r}{1 \mathrm{pc}} \right)^{-1.8}  \: {\rm \frac{GeV}  {\it c^2}  cm^{-3}}   ,
\end{equation} 
From the above two equations, we may define the core radius as
\begin{equation}
r_{c} = 16.7 \left(\frac{n_{\rm c}}{10^{14} \: {\rm cm^{-3}}} \right)^{-0.81/1.8} \: {\rm AU}.
\end{equation}
When the gas reaches the maximum of $n_c =$ 10$^{16}$ cm$^{-3}$, we have $r_c = 2$ AU and $\rho_{xc} =  5\times10^{13}$ (GeV/c$^2$) cm$^{-3}$.
At small distances $r <  r_{\rm c}$, the density varies as
\begin{equation}\label{dmcore}
\rho_{x} = \rho_{xc} \left(\frac{r}{r_c} \right)^{-0.5}.
\end{equation}

In addition to the \cite{smithetal2012} profile, we also compare with a slightly modified $r^{-2}$ DM density profile in a separate simulation, `DMA-A2'.
In this case the DM density declines as 

\begin{equation}
\rho_x = 6000 \left( \frac{r} {1 \mathrm{pc}} \right)^{-2}  \: {\rm \frac{GeV}  {\it c^2}  cm^{-3}  }     
\end{equation}

\noindent for $r > r_{c}$, but keeps a constant density within the DM core at distances $r < r_{c}$:
%$\rho_{x} \rho_{xc}$ for all $r < r_{c}$. 

\begin{equation}
\rho_x = \rho_{xc} =  6000 \left(\frac{r_{c}}{1 \mathrm{pc}} \right)^{-2}  \: {\rm \frac{GeV }  {\it c^2} cm^{-3}}     \mbox{.}
\end{equation}

%\noindent
Eqs. 16 and 17 are normalized to enclose the same DM mass as found within the central 10 pc of the original cosmological simulation.  We determined this normalization when the gas reached a maximum density of 10$^{16}$ cm$^{-3}$, which corresponds to $r_c = $2 AU and $\rho_{xc} = 6\times10^{13}$ (GeV/c$^2$) cm$^{-3}$.  Note the similarity in normalization between the DMA-A1 and DMA-A2 profiles on small scales.  

As will be further shown in Section 4.3, the Eq. 16 profile is somewhat steeper than the DM profile from our original cosmological simulation on large scales, where the DM density roughly goes as $\rho_x \propto r^{-1}$.  We are thus assuming adiabatic contraction of the DM is more effective on small scales that were unresolved in the cosmological simulation.

We run case DMA-A2 until the first sink forms, and then follow the simulation for $\sim$ 10,000 yr of sink accretion.  During the accretion stage the DM profile is kept centered upon the most massive sink.

\subsubsection{N-body Particle Profile}
We next take the simulation output of DMA-A2 at the time just prior to the formation of the first sink, and we use this as the initial state for a final simulation which includes a `live' DM halo, `DMA-L2.'  
For the `DMA-L2' case, instead of using an analytic $\rho_x$ profile we add 256$^3$ DM particles, arranged such that they have the density profile of Eqs. 16 and 17 when $r_c =$2 AU and $\rho_{xc} = 6\times10^{13}$ (GeV/c$^2$) cm$^{-3}$.  The DM distribution is thus the same as that of DMA-A2 when the DM and gas reaches maximum density. 
%(Figure \ref{dm_densprof}). 
Each DM particle has a mass of $m_{\rm DM}= 3.8\times 10^{-4}$ M$_{\odot}$, and hence the total DM mass within our cut-out box is identical to the total DM mass within the central 10 pc cube of the original cosmological simulation (6300 M$_{\odot}$).
In addition, $m_{\rm DM}$ matches the value of $m_{\rm SPH}$ to within a few tens of percent.
%Estimating $M_{\rm res,DM}$ as the total mass of $\sim$ 60 particles within a DM `smoothing kernel', 
%we find that the DM mass resolution ($M_{\rm res,DM} \sim 0.02 $ M$_{\odot}$) is similar to that of the gas 
%($M_{\rm res} \sim 0.03 $ M$_{\odot}$), allowing us to study how the DM and gas interact on similarly small scales. 
%Had we instead used only 128$^3$ DM particles, this would have required either that the mass resolution was eight times larger ($M_{\rm res,DM} \sim 0.2 $ M$_{\odot}$), or that the DM profile encompass eight times less mass ($\sim 800$ M$_{\odot}$) and thus extend out to a shorter distance of $\sim$ 1 pc.  
%%%radius (NOT diameter!) = 1 pc, M_enc = 600 M_sol
%%%radius = 1.5 = 1.5 pc, M_enc = 1000 M_sol
%DM resolution discussion
%Estimating $M_{\rm res,DM}$ as the total mass of $\sim$ 60 particles within a gravitational softening length of 3 AU, 
%we find that the DM mass resolution ($M_{\rm res,DM} \sim 0.02 $ M$_{\odot}$) is similar to that of the gas 
%($M_{\rm res} \sim 0.03 $ M$_{\odot}$), 
The low values of $m_{\rm DM}$ and gravitational softening length (3 AU) allow us to study how the DM and gas interact on similarly small scales. 
%Using only 128$^3$ DM particles would have required either that $m_{\rm DM}$ was eight times larger ($m_{\rm DM} \sim 3 \times 10^{-3} $ M$_{\odot}$), or that the DM profile encompass eight times less mass ($\sim 800$ M$_{\odot}$) and thus extend out to a shorter distance of $\sim$ 1 pc.  

As in \nocite{stacyetal2012}  Stacy et al. (2012b), we produce the DM density profiles beginning from an initial uniform
density field. This field was generated by placing particles at
glass-like positions, which was achieved by allowing randomly placed
particles to evolve under an artificial negative gravitational force
until a quasi-equilibrium configuration was reached
(\citealt{white1996}).  This method avoids small-scale fluctuations in the relative distances between particles, and it is an improvement upon a Monte Carlo sampling of the density field, which would be subject to such Poisson noise.

The particles of uniform density $\rho_0$ can then be transformed to a power law density profile
$\hat{\rho}  \propto \hat{r}^{-n}$ through the coordinate transformation $(r,\theta,\phi) \rightarrow (\hat{r},\theta,\phi)$. The new coordinates will satisfy

\begin{equation}
\hat{\rho}(\hat{r}) \hat{r}^2 {\rm sin}\theta \ud \hat{r}   \ud \theta \ud \phi =   \rho_0 r^2 {\rm sin}\theta \ud r \ud \theta \ud \phi \mbox{,}
\end{equation}

\noindent from which we can derive the relation

\begin{equation}
\hat{r} \propto r^{3/(3-n)} \mbox{, for \ }
0 \le n < 3 \mbox{.}
\end{equation}

\noindent In this way we thus acquire the desired $n=2$ profile for the live DM halo.

Similarly to the SPH particles, the N-body particles are given an adaptive gravitational softening length identical to their `smoothing length', a length which is calculated in the same way for both the SPH and N-body particles. We furthermore imposed a minimum softening length of 3 AU for the DM, the same length as the inner sink accretion radius.  In the DMA-L2 run, the DMA heating and ionization rates for gas particles with $n > 10^5$ cm$^{-3}$ were calculated based upon the $\rho_x$ value for the most nearby DM particle.  Rates for gas particles with  $n < 10^5$ cm$^{-3}$ were instead determined using Eq. 16 to reduce the numerical cost of searching for nearby DM neighbors.

\begin{figure*}
\includegraphics[width=.9\textwidth]{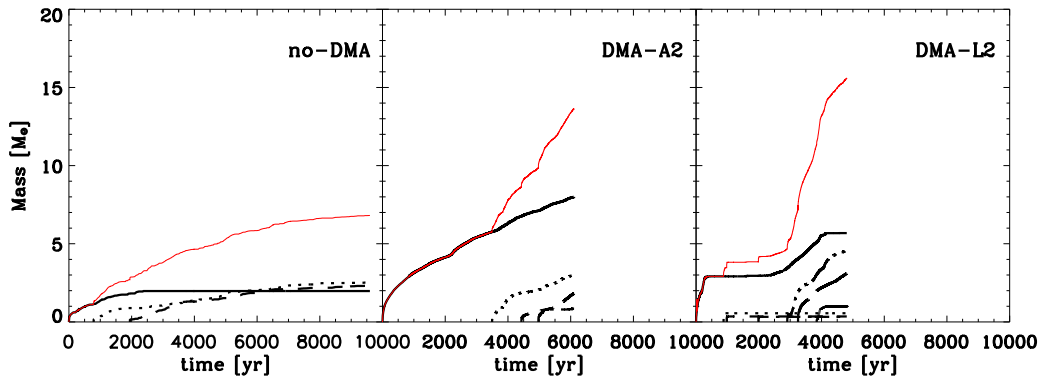}
 \caption{
 Growth of the sinks over time, as measured from the moment of initial sink formation.  
%Dashed line represents the growth of the second largest sink.
{\it Left:} Sink mass within halo with no DMA effects.  
{\it Center:} Analytic halo with $r^{-2}$ profile (DMA-A2).
{\it Right:} Live halo with evolving profile (DMA-L2).
The different black line styles represent the various sinks, while the red lines represent the total sink mass within the given simulation.
In the DMA-A2 and DMA-L2 runs, DMA effects result in more rapid gas collapse and an earlier formation time for the first sink by $\sim$ 10$^5$ yr.  However, DMA leads to a delay in the appearance of secondary sinks with respect to the formation time of the first sink, though in the DMA-L2 case it only slightly delays secondary fragmentation.  DMA additionally yields enhanced sink accretion rates. 
}
\label{sinkmass}
\end{figure*}

\begin{figure*}
\includegraphics[width=.9\textwidth]{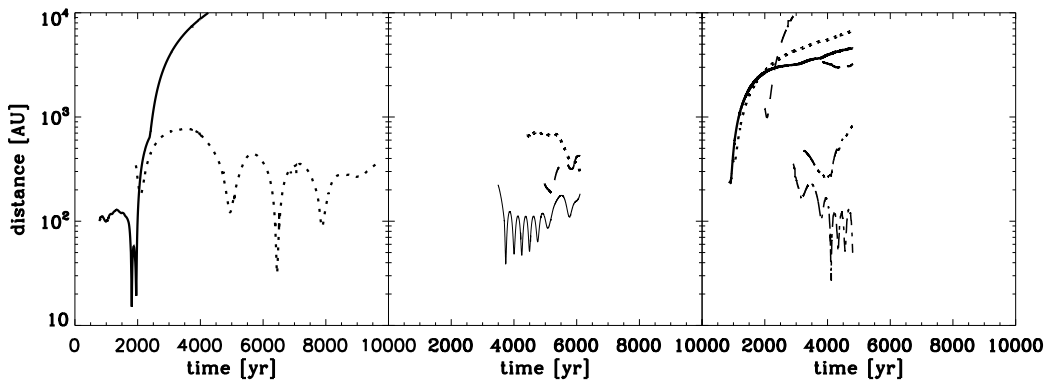}
 \caption{
 Distance of secondary sinks from the most massive sink over time.   
{\it Left:} no-DMA simulation.  
{\it Center:} DMA-A2 simulation.
{\it Right:} DMA-L2 simulation.
The most massive sink is usually the first-formed sink.  However, in the no-DMA run the second-formed sink becomes the most massive while the first-formed sink is ejected from the system, so in this case distances are measured from the second-formed sink.
The no-DMA run forms a total of three sinks.  The DMA-A2 run forms four sinks, while the DMA-L2 run has seven sinks by the end of the simulation.
}
\label{sinkdis}
\end{figure*}

\begin{figure*}
 \includegraphics[width=.95\textwidth]{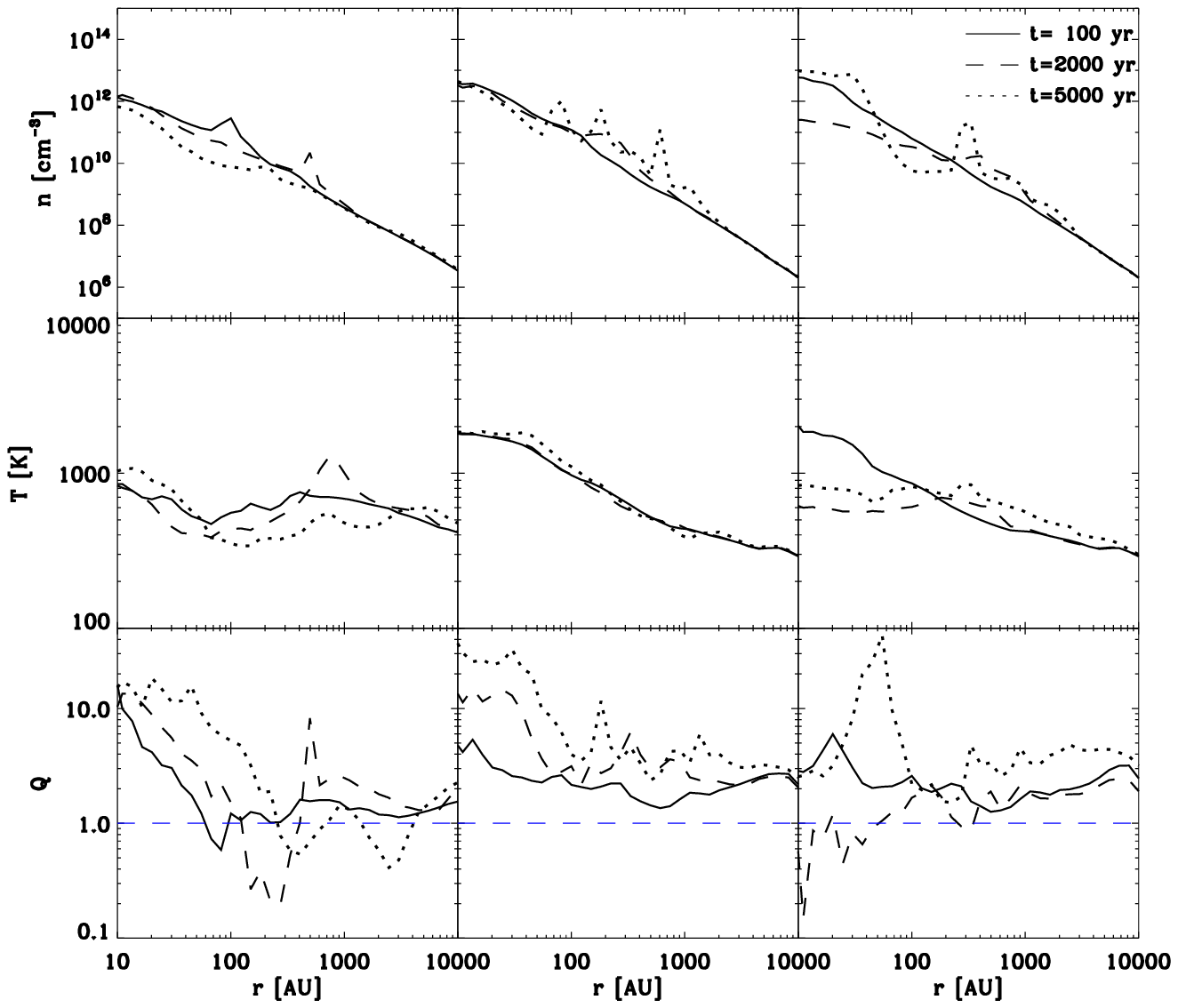}
 \caption{
Evolution of gas density, temperature, and estimated Toomre $Q$ parameter for each of the three simulations.  Profiles are measured with respect to distance from the 
the densest gas particle.
%most massive sink.
{\it Left:} no-DMA case.
{\it Center:} DMA-A2 case.
{\it Right:} DMA-L2 case.
Solid lines represent profiles taken at $t_{\rm acc} = 100$ yr, dashed lines at $t_{\rm acc} = 2000$ yr, and dotted lines at $t_{\rm acc} = 5000$ yr.  Blue dashed lines represent the minimum Toomre parameter value ($Q=1$) at which a disk can remain stable.  The lower temperatures of the no-DMA case lead to generally lower values of the Toomre parameter.  
}
\label{evol}
\end{figure*}

\subsection{Velocity Initialization}
 
The `live' DM halo (DMA-L2) case furthermore requires an initial velocity to be assigned to each DM particle.  We do this in the same way as described in \nocite{stacyetal2012}  Stacy et al. (2012b).  Given a  spherically symmetric density profile, an isotropic distribution function (DF) for the DM particles can be generated using Eddington's formula (\citealt{binney&tremaine2008}, their equation 4.46):

\begin{equation}
f(\mathcal E)=\frac{1}{\sqrt{8}\pi^2}\left[\int^{\mathcal E}_0 \frac{\ud \Psi'}{\sqrt{\mathcal E-\Psi'}}\frac{\ud^2\rho}{\ud\Psi'^2}+\frac{1}{\sqrt{\mathcal E}}\left(\frac{\ud\rho}{\ud \Psi'}\right)_{\Psi'=0}\right] 
\end{equation}
 
\noindent where $f(\mathcal E)$ is the DF in units of mass per phase space volume, e.g., g cm$^{-3}$ (cm s$^{-1}$)$^{-3}$, and $\Psi$ is the relative potential, which can be set to the negative of the gravitational potential as measured from the edge of the system.  $\rho$ is the mass density of the system at the given $\Psi$, and $\mathcal E$ is the relative energy of DM per unit mass.
If $\Psi$ and the relative energy $\mathcal E$ are known for a particle, then its velocity $v$ is given by

\begin{equation}
v = \sqrt{2(\Psi - \mathcal E)} \mbox{.}
\end{equation}

We assume the input $\rho$ profile is cut off at an outer radius of 10 pc.  We also assume the profile flattens to a uniform-density core in the central 1 AU, well inside the extent of the sink particles. 
To assign velocities to each particle, we first divided the DM profile into 3000 radial bins, centered upon the densest particle.  We then calculated the relative potential $\Psi$ at each bin.  Each particle was assigned the value for $\Psi$ corresponding to its bin, and a value for its relative energy $\mathcal E$ was randomly drawn from the DF.  The amplitude of the velocity $v$ was then found using Eq. 21.
In the appendix of \nocite{stacyetal2012}  Stacy et al. (2012b) we provide more details about randomly drawing a value of $\mathcal E$ from the distribution function $f(\mathcal E)$ .

To determine the velocity component along each Cartesian axis ($v_x$, $v_y$, and $v_z$), we next picked two angles at random, $\theta$ and $\phi$ (e.g. \citealt{widrow2000}).  Given two random numbers $p$ and $q$ ranging between 0 and 1, we set  $\theta = {\rm {cos^{-1}} (1 - 2{\it p})}$ and  $\phi = 2\pi q$.  We used these angles to map each particle in velocity space, where $v_x = v \rm \, sin\theta\, cos\phi$, $v_y = v \rm \, sin\theta\, sin\phi$, and $v_z = v \rm \, cos\theta$.    

We emphasize here that the resulting velocity distribution yields a stable DM profile. Before combining the DM particles with the gas, we evolve the DM-only profile for 10,000 yr to ensure that it undergoes minimal change (figures in Section 4.3).  It is this evolved profile that is used in simulation `DMA-L2'.  As will be further discussed in Section 4.3, it is only after interaction between the gas and DM that the DM density and velocity profile undergo substantial evolution.    

\begin{figure*}
\includegraphics[width=.8\textwidth]{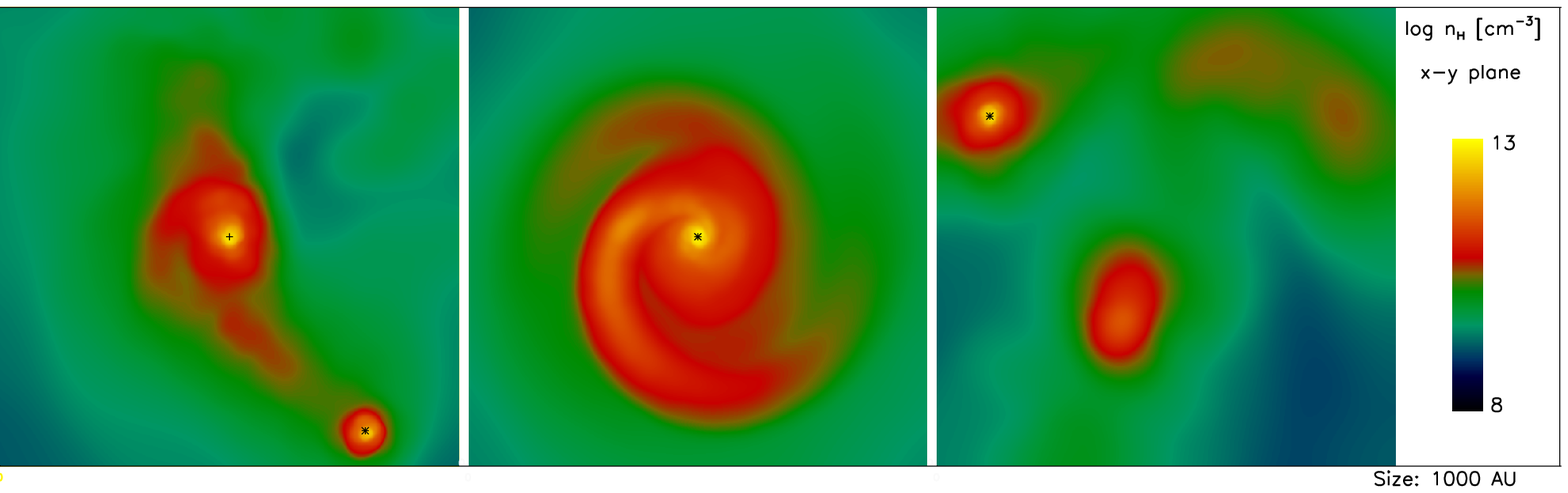}
\includegraphics[width=.8\textwidth]{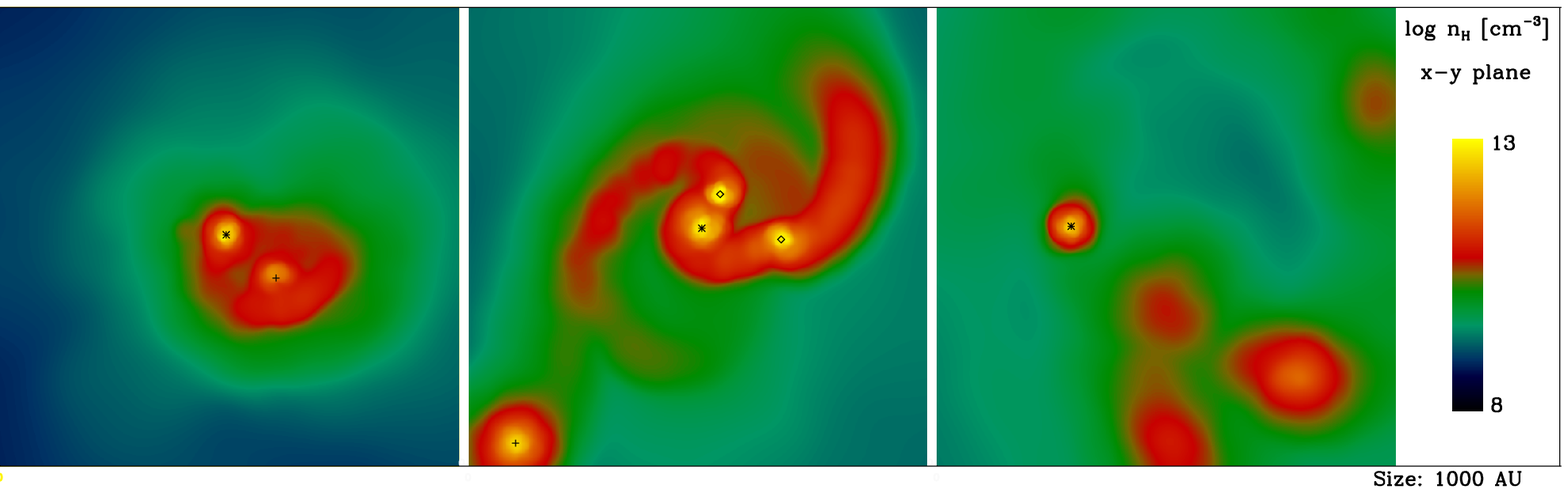}
 \caption{
 {\it Top Panels:} Morphology of gas density 2500 yr after sink formation.  Images are centered upon the center-of-mass of the sinks and gas with $n>10^{13} cm^{-3}$.  
 The panels are 1000 AU across. 
Asterisks represent the most massive sink, plus symbols depict the second-most massive sink, and diamonds represent all other sinks.
{\it Top Left:} no-DMA case.  Note that the first-formed sink does not appear because it has been ejected from the disk as a result of a three-body interaction and is $\sim$ 200 AU beyond the region shown here.
{\it Top Middle:} DMA-A2 simulation.  
The DMA-A2 run has a more massive disk with more pronounced spiral structure, and at this point fragmentation has still been suppressed due to DMA heating.
 {\it Top Right:} DMA-L2 simulation. Like the no-DMA case, the gas has undergone multiple instances of fragmentation at this point.  
{\it Bottom Panels:}  Same as top row, but at 5000 yr after initial sink formation.
 The  DMA-A2 run still has a more massive disk and pronounced spiral structure, but has formed multiple secondary sinks by this time.
 }
\label{morph_2500}
\end{figure*}

\section{Results}

In this section we first discuss the effects of DMA on the initial gas collapse and compare with Smith et al. (2012; Section 4.1). After that, we investigate how DMA affects the subsequent gaseous disk evolution, comparing the influence of the analytic and live DM profiles (Section 4.2).  Finally, we discuss the DM evolution in the case of the live DM profile (Section 4.3).

\subsection{Initial Collapse to Stellar Densities}

Our set of simulations is summarized in Table \ref{tab1}.  `Live' profiles refer to runs in which the DM structure was followed with N-body particles.  `Analytic' profiles refer to those in which a spherically symmetric DM density profile was assumed as described in Section 3.2.

The gas velocity profiles in our various cases are compared in Fig. \ref{velprof}, as determined just before the formation of the first sink particle.   
We measured the rotational and radial velocities of each gas particle with respect to the center-of-mass of the densest gas, those particles with $n > 2\times10^{11}$ cm$^{-3}$.  The  rotational velocity $v_{\rm rot}$ and  radial velocity $v_{\rm rad}$ within each logarithmically-spaced
radial bin is then taken as the mass-weighted average of the individual particle velocities within each bin.    
In a similar fashion,
we measure the turbulent Mach number $M_{\rm turb}$ over the same range of 
radial bins according to the following:
\begin{equation}
M_{\rm turb}^2 c_s^2 =\sum_{i} \frac{m_i}{M}\left(\vec{v}_i - \vec{v}_{\rm rot }  - \vec{v}_{\rm rad  }\right)^2 \mbox{,}
\end{equation}
where $c_s$ is the sound speed of the radial bin, $m_i$ is the mass of a gas particle contributing to the bin, and $M$ is the total gas mass of the bin.

\begin{figure}
\includegraphics[width=.40\textwidth]{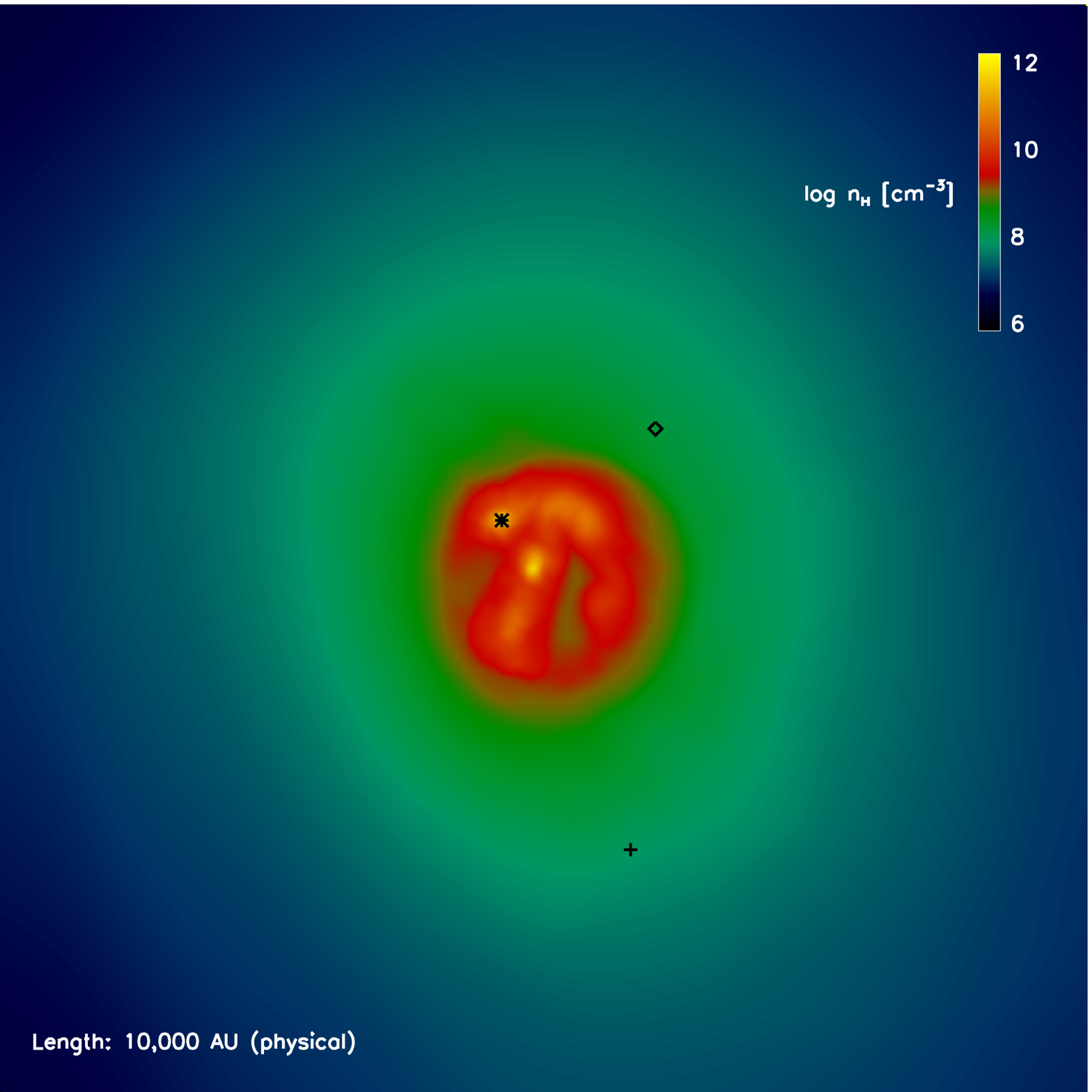}
 \caption{
%{\it Left:} 
Morphology of DMA-L2 simulation at 2500 yr after sink formation with panel size of 10,000 AU.  This simulation has four sinks at this time, though all but one have been ejected from the disk.  One of these ejected sinks is outside of the pictured region.
%{\it Right:} Same as left panel but at 5000 yr.  
 }
\label{morph_large_2500}
\end{figure}

When including DMA effects (DMA-A1 and DMA-A2, middle and right panels of Fig. \ref{velprof}), the radial infall rates are 
similar to the no-DMA case but 
slightly greater at distances beyond 1000 AU.
Rotational motion of the gas as it approaches stellar densities is more noticeably enhanced under DMA, while turbulent motion is greater at distances beyond 100 AU.  
In good agreement with \cite{smithetal2012}, we also find an increase in sound speed in the inner 1000 AU but a decrease beyond 1000 AU.  
%The free-fall velocity $v_{\rm ff}$ curves in Figure \ref{velprof}, which are based upon the enclosed mass at the given radius ($v_{\rm ff} = (G M_{\rm enc}/r)^{(1/2)}$), furthermore indicate the larger central mass at the time of sink formation in the DMA-A1 and DMA-A2 runs.  
Within the central few thousand AU, the enclosed mass $M_{\rm enc}$ is greater in the DMA cases (Fig. \ref{menc_prof}), consistent with the moderately enhanced $|v_{\rm rad}|$ at scales $> 100$ AU.
As is visible in Fig. \ref{morph}, this coincides with a more massive, extended, and flattened disk structure at this time.

As already described in detail in \cite{smithetal2012}, the DMA effects lead to a very different chemothermal evolution at densities above $\sim 10^6$ cm$^{-3}$.  
We compare the no-DMA, DMA-A1, and DMA-A2 runs in Fig. \ref{Tvsnh}.  As expected, DMA-A1 and DMA-A2 are very similar since they use similar DM density profiles.
At low densities, the gas in the DMA runs has lower temperatures than in the no-DMA case due to enhanced H$_2$ cooling rates.   The  DMA-A1 and DMA-A2 runs undergo rapid H$_2$ formation at a density of  $\sim$ 10$^{6}$ cm$^{-3}$, which is two orders of magnitude lower than the similar upturn in  H$_2$ fraction seen in the no-DMA case. This is made possible by the enhanced electron abundance at these densities, which is in turn due to DMA-induced ionizations. 
However, at densities of $\sim$ 10$^{9}$ cm$^{-3}$ the temperature undergoes a rapid increase and surpasses that of the no-DMA case.  
This coincides with the density at which the DMA-affected gas becomes fully molecular, when the combination of DMA and H$_2$ formation heating surpasses H$_2$ line cooling just sufficiently to heat the gas (see detailed discussion in section 4.2 of \citealt{smithetal2012}).    

Within the corresponding central few thousand AU, there is additionally a slight enhancement in gas density for cases with DMA as compared with the no-DMA simulation, similar to the  density enhancement at radii between $\sim$100 and 1000 AU discussed in \cite{smithetal2012}.  
This is also in accordance with the greater enclosed mass 
%and $v_{\rm ff}$ values 
seen in these same regions (Fig. \ref{menc_prof}).
 We finally note that both runs that include DMA effects show a dip in H$_2$ fraction between densities of approximately 10$^{12}$ and 10$^{13}$ cm$^{-3}$ from H$_2$ collisional dissociation.  This is due to the increased temperatures at these densities resulting from DMA-induced heating.  At the same time, H$_2$ collisional dissociation and destruction of H$_2$ through charge transfer with H$^+$ allow the gas to remain at a steady temperature of $\sim$ 2000 K (see section 4.2 and fig. 5 of \citealt{smithetal2012}).

%\begin{figure}
%\includegraphics[width=.4\textwidth]{dm_densprof.eps}
 %\caption{
% Density profile of DM particles, inserted into the cut-out box at just prior to initial sink formation. 
%This demonstrates how the interaction between the DM and rotating gas will cause the DM to gradually disperse, leading to a flattening of the density profile.
 %}
%\label{dm_densprof}
%\end{figure}

\subsection{Sink Accretion and Disk Fragmentation}

Once the first sink particles have formed, we continue the no-DMA run as well as the DMA-A2 run, where the DM peak of the analytical profile is centered on the most massive sink at all times.  
At this point, we also use the gas properties of  DMA-A2 to initialize the DMA-L2 simulation in which we add the `live' DM profile.
We see a considerable divergence in the sink accretion histories (Figs \ref{sinkmass} and \ref{sinkdis}).  
This divergence originally  stems from the differences in the initial gas collapse, 
in which DMA leads to increased H$_2$ cooling for densities less than $\sim10^9$ cm $^{-3}$.  
The lower temperatures allow DMA-affected gas to more quickly reach densities $10^9$ cm $^{-3}$, which more than compensates for the slow-down in collapse once the DMA-affected gas surpasses this density threshold and undergoes rapid heating.  
This causes the gas in the DMA runs to form the first sink $\sim 3 \times 10^5$ yr earlier, similar to \cite{smithetal2012} who found that collapse occurred $\sim10^5$ yr earlier.        
A more massive and rapidly accreting disk with more pronounced spiral structure results (Fig. \ref{morph_2500}).

\subsubsection{No-DMA and DMA-A2 cases}

In the no-DMA case, a second sink forms 600 yr after the first.  The secondary sinks are formed at 100 and 200 AU from the most massive sink (Fig. \ref{sinkdis}).
Note that the most massive sink is the first formed sink in most cases, but in the no-DMA case the
first-formed sink is ejected from the disk at $\sim$ 2000 yr, causing its mass growth to level off.  Distances in this case are thus measured from the second-formed sink, which grows to become the largest.
A total of three sinks forms in the no-DMA run, combining to only 7 M$_{\odot}$, while each individual sink reaches $\sim$ 2 M$_{\odot}$.

In contrast, in the DMA-A2 case secondary fragmentation is suppressed until 3500 yr after the initial sink formation, though by the end of the DMA-A2 run four sinks have formed.  The last sink to form in DMA-A2 can be seen as the distant clump in the bottom panels of Fig. \ref{morph_2500}.
The formation of the fourth sink at a larger distance of 600 AU indicates that fragmentation in the outer spiral arms of the disk in DMA-A2 is 
%in fact encouraged by inclusion of DMA effects.  
still possible.
This is similar to what \cite{smithetal2012} describe in their `H2' simulation.  The first sink of DMA-A2 grows at a much higher rate than that of no-DMA (Fig. \ref{sinkmass}).  This is partially because for the first 3500 yr no secondary fragment has formed yet, and the gas that would have accreted onto secondary sinks is instead available to flow onto the solitary first sink.  
However, even the combined sink mass for the no-DMA case is below that of DMA-A2 at all times. Since the DMA-A2 sinks are growing from a more massive disk, their total sink mass reaches values twice as large (14 M$_{\odot}$).

%The relative sink masses also show interesting variation between the different cases.  The sinks from the no-DMA run  

The variation in fragmentation and accretion rate can be understood by examining the disk properties in each case (Fig. \ref{evol}).  A rough estimate for whether a disk is unstable to fragmentation is given by the Toomre fragmentation criterion:

\begin{equation}
Q = \frac{c_{\rm s} \kappa}{\pi G \Sigma} < 1  \mbox{\ ,}
\end{equation}

\noindent  where $\kappa$ is the epicyclic frequency, which we set equal to the angular velocity, as appropriate for Keplerian rotation.
$\Sigma$ is the disk surface density, which we estimate as $\Sigma \sim M_{\rm enc}(r_i)/[4\pi (r_i^2 - r_{i-1}^2)]$.  $M_{\rm enc}(r_i)$ is the total mass within a cylindrical shell whose inner and outer surfaces span from $r_{i-1}$ to $r_i$ in the $x-y$ plane.  
The $x-y$ plane of each shell is oriented perpendicular to the rotational axis of the disk and centered upon the densest gas particle.  
To exclude gas at arbitrarily large heights along the shell's $z$-axis, 
the enclosed mass of each cylindrical shell includes only gas with density greater than half of the average density within the corresponding spherical shell of radius $r_i$.
The central densities of DMA-A2 are in general slightly higher than those of the no-DMA case (top row of Fig. \ref{evol}). However, the higher temperature and sound speed of the DMA-A2 case serve to stabilize the disk, just sufficiently countering the effect of the enhanced density.  

We here further examine how DMA effects yield an increased sink accretion rate.
The rate at which gas flows through the disk onto the sinks may be estimated as

\begin{equation}
 \dot{M}_{\rm disk} = 3\pi \nu \Sigma \mbox{,}
\end{equation}

\noindent where $\nu$ is estimated based upon the prescription introduced by \cite{shakura&sunyaev1973},

\begin{equation}
 \nu = \alpha_{\rm SS} H_{\rm p} c_{\rm s} \mbox{.}
\end{equation}

\noindent  Here, $H_{\rm p}$ is the pressure scale height of the disk, and $\alpha_{\rm SS}$ is a dimensionless parameter ranging between $\sim 10^{-2}$ and 1, depending on the nature of angular momentum transport in the disk.  From this we may roughly estimate that

\begin{equation}
 \dot{M}_{\rm disk} \propto c_{\rm s}  \Sigma \mbox{.}
\end{equation}
  
\noindent Changes in either disk surface density or sound speed can thus alter $\dot{M}_{\rm disk}$.

Taking the total final sink masses and total accretion times directly from the simulations, the average sink accretion rates are $\sim 2 \times 10^{-3}$ M$_{\odot}$ yr$^{-1}$ and $8 \times 10^{-4}$ M$_{\odot}$ yr$^{-1}$  in the DMA-A2 and no-DMA cases, respectively.  Both the warmer temperature (i.e. greater sound speed) and higher density in the DMA-A2 case serve to enhance the disk and sink accretion rates, though more of the variation in $\dot{M}_{\rm disk}$ is driven in particular by the difference in temperature, which is over twice as high in the DMA-A2 case within the central 100 AU.  The central temperatures translate to $c_s \simeq$ 3.8 and 2.2 km s$^{-1}$ in the DMA-A2 and no-DMA cases, respectively, thus accounting for the majority of the factor of two difference in $\dot{M}_{\rm disk}$.

\subsubsection{DMA-L2 case}

The DMA-L2 case has identical gas properties to DMA-A2 at the point of initial sink formation, but the accretion history diverges after only a few hundred years.  Its overall rate of sink growth over the first 3000 yr
%double check this later
is less than that of DMA-A2, but slightly greater than that of no-DMA.  
A second sink forms 900 yr after the first, at a distance of 200 AU from the main sink.  The DMA-L2 case thus does not exhibit the same significant delay of fragmentation as seen in DMA-A2.  It is instead more similar to the no-DMA case.  
Even when accounting for DMA-induced heating and ionization, we thus find that relatively close fragments may still form within $\sim$ 1000 yr.
The plateau in the growth of the initial sink between 500 and 2500 yr is similar to what may be expected from `fragmentation-induced starvation' (\citealt{petersetal2010a}), where the two secondary sinks intercept mass that would otherwise accrete onto the main sink.
At the same time, the interaction of the disk with the live DM particles causes the dense gas to be much more spread out and diffuse than in the other two cases (Figs. \ref{morph_2500} and \ref{morph_large_2500}). 

In the DMA-L2 case, a total of four sinks form within the first 2000 yr, and the three secondary sinks are scattered out to large distances of several thousand AU (right-hand panel of Fig. \ref{sinkdis}).  
The total sink mass undergoes a rapid increase as two additional sinks form at $\sim$ 3000 yr, similar to the increase seen in DMA-A2.  
This up-tick is more pronounced, however, since the first-formed sink accretes much more rapidly after this point as well.  
This is due to the formation of further secondaries which redistribute the angular momentum in the disk and thereby allow more mass to fall into the main sink.
By the end of the simulation the total sink mass in the DMA-L2 case is 
%check this again when sim is done
$\sim$ 16 M$_{\odot}$.  This translates to an overall average sink accretion rate of $\sim 3 \times 10^{-3}$ M$_{\odot}$ yr$^{-1}$, 
slightly more than that found in DMA-A2.
%slightly less than that found in DMA-A2.   

In the DMA-L2 simulation, the 
 early-time ($t_{\rm acc}<3000$ yr) accretion rate and the
timescale at which secondary fragmentation occurs thus fall in between the other two cases.
%, while accretion rate is the largest accretion rate of our three cases.
This follows expectation, as the DMA-L2 case begins with the same temperature enhancement as seen in DMA-A2, which initially serves to suppress fragmentation and enhance $\dot{M}_{\rm disk}$.  Then, as the central DM densities decline due to the motions of the gas (see following section), these high central temperatures are not maintained.  The addition of gravitational interaction with the DM profile, however, leads to greater amounts of density perturbation in the gas.  This in turn leads to the formation of a larger number of protostellar fragments than found in the other cases.
 
%Our DMA runs exhibit a greater amount of fragmentation than found in \cite{smithetal2012}, even after they follow the accretion for $\ga 5000$ yr, similar to the timescale of our runs. 
 %In their `H1' and `H2' runs they found no secondary fragmentation and only a single secondary sink, respectively.  
%We note that, unlike \cite{smithetal2012}, we did not include effects of heating due to protostellar accretion radiation.  In our particular case, accretion heating would have been a smaller effect due to the lower sink growth rates in our protostellar system.  While their sinks grow to $\ga$ 10 M$_{\odot}$ by the end of their runs, in our DMA-A2 run the largest sink is only 5 M$_{\odot}$ when the second sink forms, and is $\sim$ 8 M$_{\odot}$ at the end of the simulation.  
%This likely explains the greater number of sinks formed in our  DMA-A2 run as compared with  \cite{smithetal2012}.  However, they point out that their model assumed normal Pop III protostars instead of dark stars of lower effective temperature, thus giving an upper limit to the effect of accretion luminosity.  For these analytical DM profiles, the true level of fragmentation is likely somewhere in between their results and what we find in  DMA-A2.   

%accretion luminosity discussion
\subsubsection{Effect of Accretion Luminosity}

We do not include a heating term to account for the radiative feedback from the protostars.  This was due to the uncertainty in how possible DM accretion onto the protostars would affect their luminosity and radial evolution.   If a sink contains a `dark star,' then the star may have a radius of the order of 100 R$_{\odot}$. If it is instead a 5 M$_{\odot}$ MS star, then it will have a radius of $\sim$ 1 R$_{\odot}$.  
Note that the MS radius is a lower limit, while a normal 5 M$_{\odot}$ protostar more likely has a radius closer to $\sim$ 10 R$_{\odot}$ (see, e.g., discussion of primordial stellar evolution in \citealt{hosokawaetal2010}).
Typical masses and accretion rates found in our simulation are approximately 5 M$_{\odot}$ and 10$^{-3}$ M$_{\odot}$ yr$^{-1}$.  The corresponding accretion luminosity  $L_{\rm acc}$ would then range from 

\begin{equation}
L_{\rm acc} = \frac{G M_* \dot{M}}{R_*} \simeq 10^3 - 10^5 {\rm L_{\odot}} \mbox{,}
\end{equation}

\noindent where $M_*$, $\dot{M}$, and $R_*$ are the mass, accretion rate, and radius of the star. 
As described in \cite{smithetal2011}, the resulting accretion luminosity heating rate can be estimated as 

\begin{equation}
\Gamma_{\rm acc} = \rho \kappa_P \frac{L_{\rm acc}}{4 \pi r^2} \, {\rm erg \, cm^{-3} \, s^{-1}}
\end{equation}

\noindent where $\rho$ is the gas density, $\kappa_P$ is the Planck mean opacity, and $r$ is the distance of the gas from the emitting star.  

While we omit $L_{\rm acc}$ for simplicity, we may estimate its effect relative to DMA-induced heating.
At a radius of 100 AU, $\rho \sim 10^{-13}$ g cm$^{-3}$ and $\kappa_P \sim 10^{-6}$ cm$^{2}$ g$^{-1}$ (see \citealt{mayer&duschl2005}).  This yields 
$\Gamma_{\rm acc} = 10^{-14}  - 10^{-12}$ erg cm$^{-3}$ s$^{-1}$.
In comparison, the DMA heating rate for the DM profile taken from \nocite{smithetal2012} Smith et al. (2012, DMA-A1 case) is 10$^{-10}$ erg cm$^{-3}$ s$^{-1}$.
For the stars in our simulation, this estimate shows that we can generally expect DMA heating to dominate over accretion luminosity heating while the DM profile maintains high central densities.  When the central DM density declines in the DMA-L2 run, however, $\Gamma_{\rm acc}$ may again become dominant, and the fragmentation seen at later times may be overestimated.  This similarly applies to our no-DMA run.  

Indeed, our DMA runs exhibit a greater amount of fragmentation than found in \cite{smithetal2012}, whose simulations did account for protostellar accretion radiation.  They follow the accretion for $\ga 5000$ yr, similar to the timescale of our runs. 
 In their `H1' and `H2' runs they found no secondary fragmentation and only a single secondary sink, respectively.  
%We note that, unlike \cite{smithetal2012}, we did not include effects of heating due to protostellar accretion radiation.  
In our particular case, accretion heating would have been a smaller effect due to the lower sink growth rates in our protostellar system.  While their sinks grow to $\ga$ 10 M$_{\odot}$ by the end of their runs, in our DMA-A2 run the largest sink is only 5 M$_{\odot}$ when the second sink forms, and is $\sim$ 8 M$_{\odot}$ at the end of the simulation.  
This likely explains the greater number of sinks formed in our  DMA-A2 run as compared with  \cite{smithetal2012}.  However, they point out that their model assumed normal Pop III protostars instead of dark stars of lower effective temperature, thus giving an upper limit to the effect of accretion luminosity.  For these analytical DM profiles, the true level of fragmentation is likely somewhere in between their results and what we find in  DMA-A2.  It is still the case that at late times, while fragmentation may be suppressed by protostellar feedback effects, it will not be suppressed through DMA-induced heating and ionization.  

\begin{figure}
\includegraphics[width=.45\textwidth]{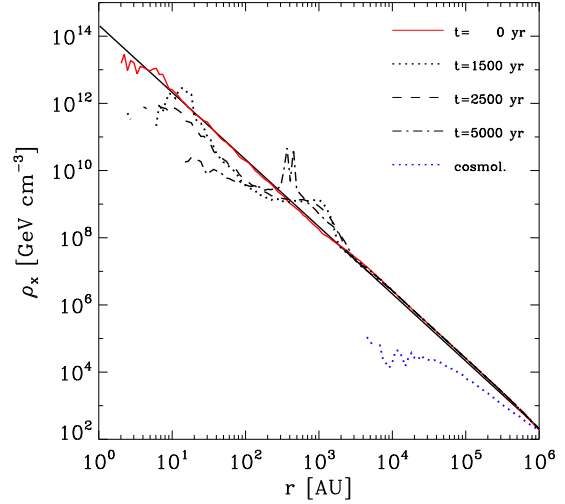}
 \caption{
DM 
%and gas 
evolution in the DMA-L2 simulation.
%{\it Upper Left:}  
DM density profile, with distance taken relative to 
the densest gas particle.
%the most massive sink.  
Black line is the $r^{-2}$ profile to which the DM particles were first initialized.
Before the gas and DM profiles were combined, 
the DM alone was evolved for 10,000 yr and showed little change (solid red line).  
The solid red line is thus the evolved DM profile that is combined with the gas at the point of initial sink formation to initialize DMA-L2 ($t=0$ yr).
Dotted line shows how this profile has evolved by 1500 yr, while dashed line shows the profile at 2500 yr and dash-dotted line is for 5000 yr.
Blue dotted line shows the DM profile from the original cosmological simulation.
The live DM profiles we generated for our refined DMA-L2 case assumed significant adiabatic contraction on small scales which were unresolved in the cosmological simulation. The mutual interaction between gas and DM causes significant decline in the central DM densities. 
%{Top right and bottom panels:} 
%Profile of the gas at 0 yr (solid black line) and 2500 yr (red line).  The peaks between 1000 and 10,000 AU represent locations of the secondary sinks.  Note the decline in density within the central $\sim$ 1000 AU and the corresponding drop in temperature.  This is due to the dispersion of the central star-forming disk.  
  }
\label{dm_evol}
\end{figure}

\begin{figure}
\includegraphics[width=.45\textwidth]{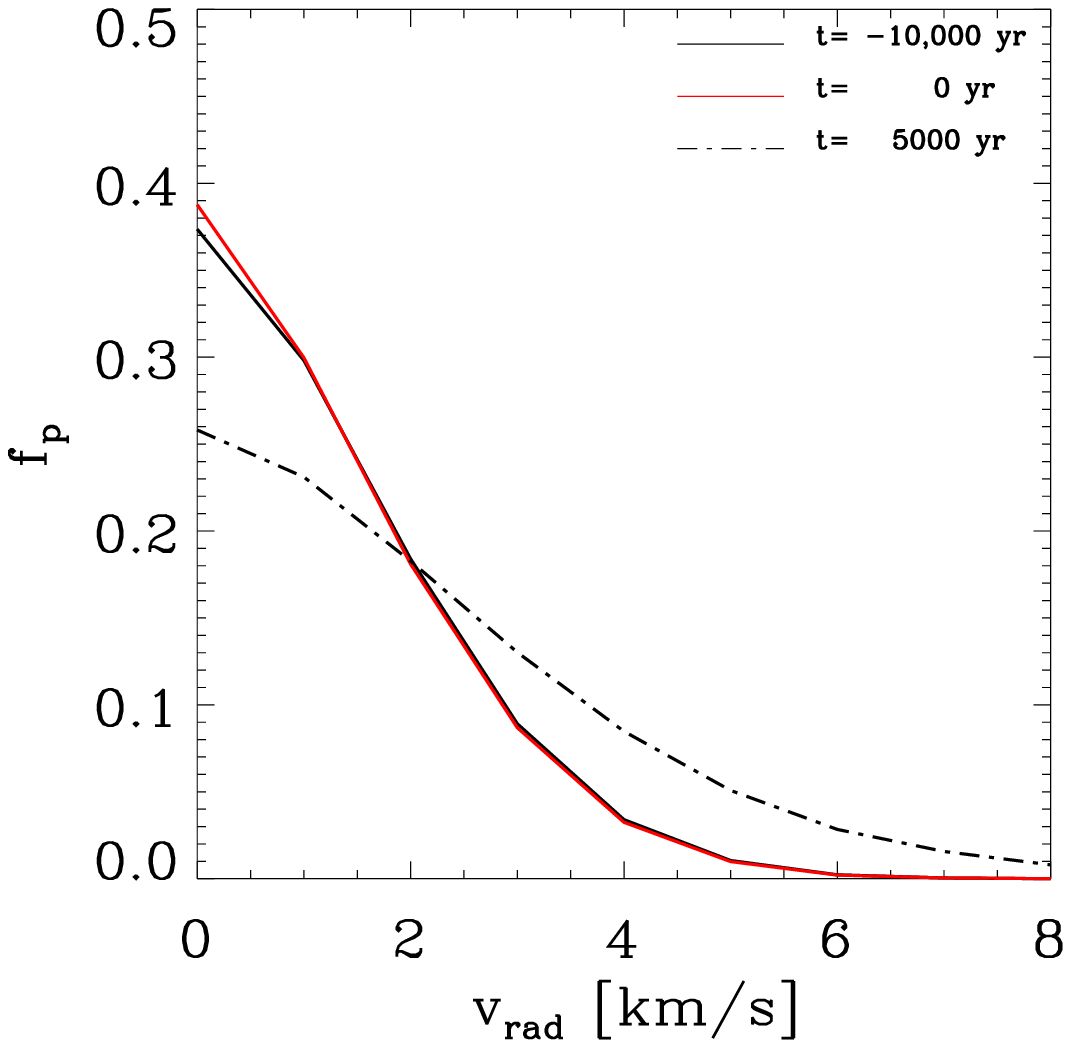}
\includegraphics[width=.45\textwidth]{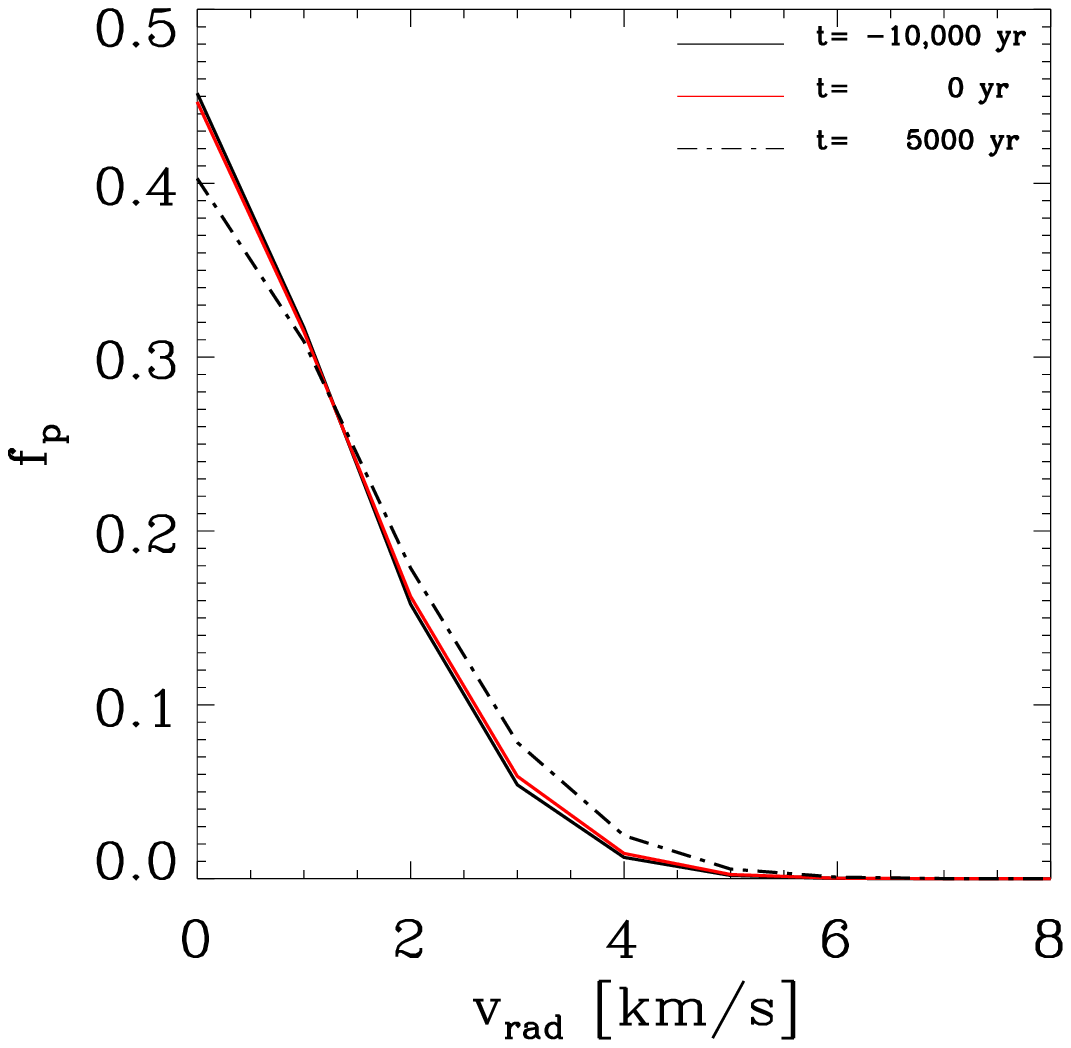}
\caption
{
{\it Top} Velocity distribution of the DM within 10$^4$ AU of the densest DM particle.
{\it Bottom} Distribution of the DM beyond 10$^4$ AU of the densest DM particle.
Black line is the velocity distribution at the point the DM is initialized.  
Red line is the distribution after the DM alone is evolved for 10,000 yr.  
This evolved distribution (red line) is used at t=0 yr for the DMA-L2 simulation.
Dash-dotted line is the distribution at the end of the DMA-L2 simulation.
The distribution changes little during the 10,000 yr of DM-only evolution.  
After the addition of the gas for DMA-L2 simulation, 
the distribution shows significant change in only 5000 yr, particularly in the central 10$^4$ AU.  
}
\label{dm_velprof}
\end{figure}

\subsection{Evolution of DM Profiles}

Fig. \ref{dm_evol} illustrates the gradual evolution of the central DM density in the DMA-L2 run, where at 1500 yr the central $\sim$ 400 AU has decreased in density.  
The DMA-L2 run has an enclosed gas mass of $\sim$ 5 M$_{\odot}$ and 20 M$_{\odot}$ at distances of 100 and 1000 AU.
For gas undergoing roughly Keplerian rotation around the main sink, where $v_{\rm Kep} \sim \sqrt{G M_{\rm enc,gas}/r}$, typical rotation times are:

\begin{equation}
t_{\rm rot} = \sqrt{ \frac{r^3} {G \: M_{\rm enc,gas}(r)} } \mbox{,}
\end{equation}

\noindent  where $M_{\rm enc,gas}(r)$ is the enclosed gas mass at distance $r$ from the sink.
At distances of 100 and 1000 AU, this corresponds to $t_{\rm rot} \sim$ 100 and 1000 yr, roughly consistent with the timescales for the density decline seen in DMA-L2.  This lends evidence that the rotational motion of the disk is indeed the cause of the decline of the central DM profile.

In addition, we compare the gas kinetic energy with the DM gravitational energy: 

\begin{equation}
K_{\rm gas}(r) \sim \frac{1}{2} \: M_{\rm enc,gas}(r) \: v_{\rm Kep}^2(r) \mbox{,}
\end{equation}

\noindent where $v_{\rm Kep}(r)$ is the Keplerian velocity at the given radius $r$, and

\begin{equation}
U_{\rm DM}(r) \sim \frac{G \: M_{\rm enc,DM}^2(r)} {r} \mbox{.}
\end{equation}

\noindent For $r=1000$ AU, $M_{\rm enc,gas} \sim 20$ M$_{\odot}$ and $v_{\rm Kep} \sim$ 4 km s$^{-1}$, so $K_{\rm gas} \sim 3 \times 10^{45}$ erg.
In comparison, $M_{\rm enc,DM} \sim 5$ M$_{\odot}$ and $U_{\rm DM} \sim 4 \times 10^{44}$ erg, almost one order of magnitude below the kinetic energy of the gas.  The gas motion is thus sufficiently energetic to disperse the DM.

We further support our claim that the DM profile evolution is not due to numerical effects, but to the physical effect of the gravitational interaction between the clumpy baryon and DM components, by confirming
that the DM profile remains stable in the absence of gas.  
We verified this through a test simulation in which we evolved the gas-free live DM halo for 10,000 yr, finding minimal change in the density and velocity profiles (Figs. \ref{dm_evol} and \ref{dm_velprof}; see also section 3.1 in \nocite{stacyetal2012}  Stacy et al. 2012b). 
%scattering effects discussion
In contrast, once the gaseous disk begins to act on the DM, the density and velocity profiles rapidly evolve, particularly in the central regions.
Thus, despite the limited mass resolution of the DM particles within our simulation, spurious effects from DM-DM two-body scattering does not significantly affect the DM profiles.  
%However, we note that two-body relaxation does become problematic when we do not apply the minimum softening length of 1 AU. 
In the related study of \nocite{stacyetal2012} Stacy et al. 2012b, which also followed the evolution of DM `live' profiles under the influence of a gaseous disk, a range of softening lengths from 5 to 50 AU was employed.  This study found a similar flattening of the central DM profile over equivalent timescales under this range of resolution lengths, confirming that this effect is not resolution-dependent.

The timescale for two-body relaxation for a system of $N$ bodies with mass $m$ can be estimated as

\begin{equation}
t_{\rm rel} \sim t_{\rm cr} 0.1 N / {\rm ln} N \mbox{,}
\end{equation}
where $t_{\rm rel}$ is the relaxation time, and $t_{\rm cr}$ the crossing time (\citealt{binney&tremaine2008}).  
If we consider the central 1000 AU, this includes approximately 5 M$_{\odot}$ of DM, or $N \sim$ 10$^4$ particles. For $t_{\rm cr} \sim 1000 \,{\rm AU} / 5 \, {\rm km \, s^{-1}} \sim 3 \times 10^{10}$ s, this results in $t_{\rm rel} \sim 3 \times 10^{12}$ s or $10^5$ yr.  Thus, as expected from the above numerical tests, $t_{\rm rel}$ is over an order of magnitude too long for two-body relaxation to take effect within our simulation time.  
Including the SPH particles, which have similar mass to that of the DM particles, further increases this estimate.  
For instance, even on smaller scales of 100 AU, the total number of SPH and DM particles is $N \sim 10^4$, and $t_{\rm rel} \sim 10^4$ yr.  However, in the simulation the evolution of the DM profile on these scales happens in a fraction of this time.

The limited mass resolution of our DM particles may also generate unphysical dynamical friction upon the DM.
For a system of bodies with mass $m$ and a Maxwellian velocity distribution, the acceleration due to dynamical friction on a body of mass $M$ travelling through this system is:

\begin{equation}
\frac{d{\bld v}_M}{dt} \simeq \frac{4 \pi {\rm ln} \Lambda \left(M + m\right) \rho}{v_M^3} \left[{\rm erf}(X) - \frac{2X}{\sqrt{\pi}}e^{-X^2} \right ] {\bld v}_M \mbox{.}
\end{equation}

\noindent (\citealt{binney&tremaine2008}).  $X=v_M/\sqrt{2}\sigma$, $\sigma$ is the velocity dispersion, 
$v_M$ is the velocity of the particle with mass $M$, 
$\rm{ln} \Lambda$ is the Coulomb logarithm, and $\rho$ is the density of the system.  We approximate $v_M \sim 5$ km s$^{-1}$, the function in brackets to be of order one, $\rho \sim 10^{-15}$ g cm$^{-3}$, $M=m=5\times10^{-4}$ M$_{\odot}$, and the Coulomb logarithm to be 

\begin{equation}
{\rm ln} \Lambda  \equiv {\rm ln} \left[\frac{b_{\rm max} v_m^2}{G(M+m)}\right] \sim \rm{ln} 100 \sim 5 \mbox{.} 
\end{equation}

\noindent Note that we have approximated the impact parameter $b_{\rm max}$ to be of order the gravitational softening length of 3 AU.  We can then write a simplified form of Equation (33):

\begin{equation}
\frac{d{\bld v}_M}{dt} \simeq \frac{20 \pi  \left(M + m\right) \rho}{v_M^2}  \mbox{.}
\end{equation}

\noindent  This yields $d{\bld v}/dt \sim 4 \times 10^{-14}$ km s$^{-2}$, or a change of 0.01 km s$^{-1}$ over a period of 10$^4$ yr.  Dynamical friction between SPH and DM particles is thus unlikely to be significant.  On the other hand, dynamical friction between a 5 M$_{\odot}$ sink and the DM particles may be orders of magnitude more significant.  However, this is then a truly physical effect dominated by the mass of the sink, which is much larger than both the WIMP particle mass and the mass of our simulated DM particles.

It is thus through mutual physical gravitational scattering between the gas and DM that gravitational accretion of DM into the densest gaseous regions halts within $\sim 1000$ yr after the first sink forms.  The DM density instead declines in most of the central 1000 AU.  This reduces the effect of DMA on the protostellar disk, and it furthermore prevents the DM from maintaining a dark star through `scattering accretion'.   Previous studies (e.g. \nocite{freeseetal2008b} Freese et al. 2008b, \citealt{yoonetal2008,ioccoetal2008, spolyaretal2009}) have suggested that, even without steady gravitational accretion, dark stars may survive indefinitely if the surrounding DM medium remains sufficiently dense, usually $\rho_x > 10^{10} - 10^{11}$ GeV/c$^2$ cm$^{-3}$.  In this process of scattering accretion, a star may capture a DM particle if the DM scatters off the condensed stellar gas.  While \cite{sivertsson&gondolo2011} found that scattering accretion will continue for less than approximately 10$^5$ yr, we find that $\rho_x$ is sufficiently reduced within $\sim$ 5000 yr for scattering accretion to halt within the majority of the protostellar disk.  

We note the possible exception of the region around two of the seven sinks which have formed by the end of the DMA-L2 simulation (see the two peaks around 40 AU in $t=5000$ yr line of Fig \ref{dm_evol}). Adiabatic contraction around these two later-forming sinks has formed two DM density peaks with $\rho_x \sim 10^{11}$ GeV/c$^2$ cm$^{-3}$.  However, with sufficient time these peaks are also likely to decline, and the majority of the sinks are in a DM medium with densities lower than this.

The speed at which the DM cusp is transformed will vary depending on the particular DM profile as well as the properties of the central gas.  Our DM profile was initialized to have an enclosed mass that was consistent with the original cosmological simulation (blue dotted line in Fig. \ref{dm_evol}).  However,  we note that at any given radius our $\rho_x$ value is a factor of several below the analytic adiabatic contraction results found by \cite{spolyaretal2008} and used in Smith et al. (2012; see also discussion of DM adiabatic contraction in \citealt{blumenthaletal1986}).  This points to the uncertainties in the formation and evolution of DM profiles on small scales and the discrepancies between analytic DM models, numerical simulations, and observations (e.g., the well-known `cusp versus core' problem; \citealt{moore1994,burkert1995,debloketal2001,gentileetal2005,spekkensetal2005,battagliaetal2008}).  Even for a minihalo which undergoes more efficient DM adiabatic contraction than that found in  our simulation, however, the dispersal of the central DM will most likely be delayed but not prevented.  As gas continues to condense towards the center of the minihalo, sufficient kinetic energy will inevitably build up to transform the DM cusp to a core.  Future simulations will cover a wider range of possible DM profiles.  

\section{Summary and Conclusions}

We have performed a series of simulations to explore the effect of DMA-induced ionization and heating on the formation of Pop III stellar systems.  
Our study included the first simulation to model both DMA heating and ionization rates as well as the gravitational interaction between the primordial gas and DM.
In agreement with \cite{smithetal2012}, who used a static DM profile and ignored the mutual interaction of DM and gas, 
we find an enhanced H$_2$ abundance and more rapid cooling as the gas collapses to densities of 10$^9$ cm$^{-3}$, caused by the DMA-induced ionization which increases the fractional abundance of electrons that may catalyze H$_2$ formation.  
At densities greater than 10$^9$ cm$^{-3}$, DMA-induced heating leads to warmer gas temperatures which stabilize the accretion disk.
The formation of secondary protostellar fragments is thus delayed by $\sim 3000$ yr when we employ a non-evolving analytical DM density profile. In addition, the total stellar system accretes at a more rapid rate under the effects of DMA.  Due to our omission of the effects of protostellar heating from accretion luminosity, the suppression of fragmentation was not as strong as in \cite{smithetal2012}.  
%However, our protostellar masses were also somewhat lower than theirs, so in our case the protostellar heating is unlikely to be as dominant an effect.

We find that these DMA effects are mitigated when we represent the DM with highly-resolved N-body particles, allowing us to follow the evolution of the DM density profile due to gravitational interaction with the gas.  The central DM density is gradually reduced as the motion of the gas scatters the DM to increasingly large distances from the peak.  At the same time, the formation of secondary protostellar fragments is delayed by only a few hundred years.  The overall sink accretion rate is
%double check this
more rapid than when DMA effects are ignored, but similar to the accretion rate found when using an analytic DM profile. Surprisingly, the number of secondary protostars formed was highest in this `live' DM simulation.  Along with the influence of DM through DMA-induced heating and ionization, the mutual gravitational interaction between the DM and gas caused the gas disk to be more diffuse and extended than when using an analytic DM profile.  This leads to enhanced levels of perturbation in the disk and the formation of six secondary protostars, as compared to two and three secondaries in the case of no DM or an analytic DM profile.

As pointed out in \cite{smithetal2012} and also discussed in \cite{gondoloetal2013}, neither our simulations nor those of \cite{smithetal2012} follow the gas collapse to protostellar densities.  To determine whether an equilibrium dark star forms, the gas collapse must be followed to higher densities of $10^{17}$ cm$^{-3}$.  \cite{gondoloetal2013} argue that such a dark star may indeed still form within the unresolved sinks of  \cite{smithetal2012}, and that the DMA suppression of fragmentation may allow the peaked DM profile to remain undisturbed.  However, our DMA-L2 case shows that a live DM profile will not maintain the necessary high central densities,  even when accounting for the stabilizing effects of DMA on the gas (Fig. \ref{dm_evol}).  The strong interaction between not only the DM and sinks, but also between the DM and gas, is a strong implication that even the motion of a disk around a single star will affect the DM density profile in similar ways to what we have presented here.  
Therefore, DMA likely did not have long-term effects on the fragmentation of primordial gas and the degree of multiplicity among Pop III stars.  
Dark stars, as initially envisioned, may thus not have existed, since the required high DM densities did not persist for sufficiently long times, as shown here. However, within plausible WIMP scenarios, DM annihilation heating would nevertheless be important in modifying the thermodynamics of primordial gas, at least for part of its evolution.
The corresponding heating and ionization terms should therefore be
included in more complete numerical work.
Future simulations, which include DMA effects as well as processes such as protostellar feedback and magnetic fields, will continue to add clarity to our understanding of Pop~III stars.

\section*{Acknowledgments}
A.S. is grateful to John Mather for helpful comments and discussion.
The authors thank Simon Glover for helpful comments.
Resources supporting this work were provided by the NASA High-End Computing (HEC) Program through the NASA Advanced Supercomputing (NAS) Division at Ames Research Center and the Texas Advanced Computing Center (TACC). 
A.H.P. receives funding from the European Union's Seventh Framework Programme (FP7/2007-2013) under grant agreement number 301096-proFeSsoR.
V.B. acknowledges support from NSF grant
AST-1009928 and NASA ATFP grant
NNX09AJ33G.

\bibliographystyle{mn2e}
\bibliography{dma}{}

\label{lastpage}

\end{document}